\documentclass[aps,showpacs,superscriptaddress,nofootinbib,amsmath,amssymb]{revtex4}
 \usepackage{graphicx}
\usepackage{dcolumn}
\usepackage{bm}
\usepackage{lipsum}
\usepackage{adjustbox}
\usepackage{multirow}
\usepackage{amsmath}
\usepackage{amssymb}
\usepackage{enumerate}

\graphicspath{./fig}
\begin{document}
 \title{Weak structure functions in $\nu_l-N$ and $\nu_l-A$ scattering with nonperturbative and higher order perturbative QCD effects}
\author{F. Zaidi}
\affiliation{Department of Physics, Aligarh Muslim University, Aligarh - 202002, India}
\author{H. Haider\footnote{on leave from Aligarh Muslim University}}
\affiliation{Fermi National Accelerator Laboratory, Batavia, Illinois
60510, USA}
\author{M. Sajjad Athar\footnote{Corresponding author: sajathar@gmail.com}}
\author{S. K. Singh}
\affiliation{Department of Physics, Aligarh Muslim University, Aligarh - 202002, India}
\author{I. \surname{Ruiz Simo}}
\affiliation{Departamento de F\'{\i}sica At\'omica, Molecular y Nuclear,
and Instituto de F\'{\i}sica Te\'orica y Computacional Carlos I,
Universidad de Granada, Granada 18071, Spain}
\begin{abstract}
We study the effect of various perturbative and nonperturbative QCD corrections on the free nucleon structure functions
($F_{iN}^{WI}(x,Q^2); ~i=1-3$) and their implications in the determination of nuclear structure functions. The evaluation of the nucleon structure functions has
been performed by using the MMHT 2014 PDFs parameterization, and 
the TMC and HT effects are incorporated following the works of Schienbein et al. and Dasgupta et al., respectively.
These nucleon structure functions are taken as input in the determination of nuclear structure functions. The 
numerical calculations for the $\nu_l/\bar\nu_l-A$ DIS process have been performed by 
incorporating the nuclear medium effects like Fermi motion, binding energy, nucleon correlations, mesonic contributions, shadowing and antishadowing in several nuclear targets such as carbon, 
polystyrene scintillator, iron and lead which are being used in MINERvA, and in argon nucleus which is relevant for the ArgoNeuT and DUNE experiments. 
The differential scattering cross sections $\frac{d^2\sigma_A^{WI}}{dx dy}$ and 
$(\frac{d\sigma_A^{WI}}{dx}/\frac{d\sigma_{CH}^{WI}}{dx})$ have also been studied in the kinematic region of MINERvA experiment. The theoretical results are compared with the recent experimental data of MINERvA and the earlier data 
of NuTeV, CCFR, CDHSW and CHORUS collaborations. Moreover, a comparative analysis of the present results for the ratio $(\frac{d\sigma_A^{WI}}{dx}/\frac{d\sigma_{CH}^{WI}}{dx})$, and the results
from the MC generator GENIE and other phenomenological models of Bodek and Yang, and Cloet et al., has been performed in the context of MINERvA experiment. The predictions have also been made 
for $\bar\nu_l-A$ cross section relevant for MINERvA experiment.

  \end{abstract}
\maketitle
 
 \section{Introduction} 

The physicists are making continuous efforts both in the theoretical as well as experimental fields for a better understanding of hadronic structure and parton dynamics of nucleons, in a wide
range of energy ($E$) and momentum transfer square ($Q^2$). The deep inelastic scattering process
 with large values of four momentum transfer square has been used for a long time to explore the partonic distribution in the nucleon. Therefore, several studies
 are available concerning the perturbative region of high $Q^2$, however, much emphasis has not been given to the nonperturbative region of low $Q^2$. 
In a recent theoretical work~\cite{Zaidi:2019mfd}, we have emphasized the effects of perturbative and nonperturbative QCD corrections in the evaluation of electromagnetic nucleon and 
 nuclear structure functions. In the present paper, we have extended our analysis to the weak sector by considering the QCD corrections in the charged current (anti)neutrino induced deep 
 inelastic scattering (DIS) process off free nucleon and nuclear targets. This study is to understand the effects of nonperturbative corrections such as target mass correction (TMC) 
 and higher twist (HT) effects, 
 perturbative evolution of parton densities, nuclear medium modifications, isoscalarity corrections and the center of mass (CoM) energy cut on the weak nuclear 
 structure functions. Using these nuclear structure functions,
 the scattering cross section has been determined. This study is relevant for the development of precision experiments in order to determine accurately neutrino oscillation parameters,
determination of mass hierarchy in the neutrino sector, etc., besides the intrinsic interest of understanding nucleon dynamics in the nuclear medium. For example, the planned DUNE
experiment at the Fermilab~\cite{Abi:2018alz, Acciarri:2015uup} is expected to get more than $50\%$ contribution to the event rates from 
the intermediate region of DIS and resonance production processes from nuclear targets. The ArgoNeuT collaboration~\cite{Acciarri:2014isz} has also measured the inclusive
$\nu_l/\bar\nu_l-^{40}Ar$ scattering cross section in the low energy mode.

The ongoing MINERvA experiment at the Fermilab is using intermediate energy (anti) neutrino beam, with the average energy of $\sim$6 GeV, where significant events contribute from the DIS processes.
 MINERvA has measured the scattering cross sections on the different nuclear targets ($^{12}$C, CH, $^{56}$Fe and $^{208}$Pb) in 
 the energy region, where various reaction channels such as quasielastic scattering (QES), inelastic scattering (IES) and DIS contribute, and reported the ratio of charged current deep inelastic differential 
 scattering cross sections i.e., $\frac{d\sigma^{C}/dx}{d\sigma^{CH}/dx}$, $\frac{d\sigma^{Fe}/dx}{d\sigma^{CH}/dx}$ and $\frac{d\sigma^{Pb}/dx}{d\sigma^{CH}/dx}$~\cite{Mousseau:2016snl}. For the DIS, the
 results have been analyzed by applying a cut on the four momentum transfer square $Q^2\ge 1$ GeV$^2$ and the center of mass energy $W\ge2$ GeV, for the neutrino induced processes and their analysis 
 is going on for the antineutrino induced channel. 
 They have compared the observed results with the phenomenological models like those being used in GENIE 
 Monte Carlo (MC) neutrino event generator~\cite{Andreopoulos:2009rq}, Bodek-Yang modified phenomenological parameterization~\cite{Bodek:2010km} as well as from the phenomenological study of Cloet et al.~\cite{Cloet:2006bq}.
 It may be observed from the MINERvA analysis~\cite{Mousseau:2016snl} that there is large variation ($\sim 20\%$) when all the three phenomenological studies are compared. Furthermore, it is important to point out that in the MC
 event generators, the DIS cross sections are
 extrapolated phenomenologically to the region of low $Q^2$ in order to obtain the neutrino event rates. In this region, there is lack of agreement between the experimental results from MINERvA and 
 the results obtained from the various phenomenological analyses.

Therefore, it is important to understand nuclear medium effects
 specially in the low $Q^2$ region (1-5 GeV$^2$) in order to reduce the systematics, in the neutrino oscillation analysis which contributes $\sim 25\%$ uncertainty to
 the systematics. The DIS cross section is described in terms of the nucleon structure functions, for example, by using $F_{1N}(x,Q^2)$ 
and $F_{2N}(x,Q^2)$ in the case of electromagnetic interaction while for the weak interaction there is one more structure function $F_{3N}(x,Q^2)$, that arises due to the parity violation.
 In the kinematic region of $Q^2 \to \infty,~\nu \to \infty$, such that $x={Q^2 \over 2M_N\nu} \to$constant,
the nucleon structure functions become the function of dimensionless variable $x$ only, and $F_{1N} (x) $ and $F_{2N} (x)$ satisfy the Callan-Gross relation~\cite{Callan:1969uq}:
\begin{eqnarray}\label{cg}
   F_{2N} (x) &=& 2 x F_{1N} (x).
\end{eqnarray}
It implies that the Callan-Gross relation enables us to express the $\nu_l-N$ scattering cross section, in the massless limit of lepton, in terms of only two nucleon structure 
functions $F_{2N} (x)$ and $F_{3N} (x)$. Through the explicit
evaluation of the nucleon structure functions, one may write them in terms of the parton distribution functions (PDFs) which provide information about the momentum distribution of partons 
within the nucleon. Presently, various phenomenological parameterizations are available for the free nucleon PDFs. The different phenomenological groups have also proposed the nuclear PDFs
which are not a simple combination of free proton and free neutron PDFs. 
 In the phenomenological analyses 
 the general approach is that the nuclear PDFs are obtained using the charged lepton-nucleus 
 scattering data and the ratios 
 of the structure functions e.g. $\frac{F_{2A}}{F_{2A^\prime}}$,  $\frac{F_{2A}}{F_{2D}}$ are analyzed, where $A, A^\prime$ represent any two nuclei and $D$ stands for the 
 deuteron, to take into account the nuclear correction factor. 
 While determining the nuclear correction factor, the information regarding nuclear modification is also utilized from the Drell-Yan cross section ratio like  
 $\frac{\sigma_{pA}^{DY}}{\sigma_{pD}^{DY}}$, $\frac{\sigma_{pA}^{DY}}{\sigma_{p{A^\prime}}^{DY}}$, where $p$ stands for proton beam. Furthermore, the information
 about the nuclear correction factor is also supplemented
 by high energy reaction data from the experiments at LHC, RHIC, etc.
 This approach has been used by Hirai et al.~\cite{Hirai:2007sx},
Eskola et al.~\cite{Eskola:2009uj}, Bodek and Yang~\cite{Bodek:2010km}, 
 de Florian and Sassot~\cite{deFlorian:2011fp} and others.
The same nuclear correction factor is taken for the weak DIS processes. For example, Bodek and Yang~\cite{Bodek:2010km} have obtained the nuclear correction
factors for carbon, iron, gold and lead using the 
charged lepton DIS data and applied the same nuclear correction factor to calculate the weak structure functions $2 x F_{1A}^{WI}(x,Q^2)$, $F_{2A}^{WI}(x,Q^2)$ and $x F_{3A}^{WI}(x,Q^2)$. de Florian
et al.~\cite{deFlorian:2011fp} have analyzed $\nu_l-A$ DIS data, the charged lepton-nucleus 
 scattering data and Drell-Yan data to determine the nuclear corrections due to the medium effects. Their~\cite{deFlorian:2011fp} conclusion is that the same nuclear correction factor can describe the 
 nuclear medium effect in $l^{\pm}-A$ and $\nu_l-A$ DIS processes.  
 In the other approach nuclear PDFs are directly
 parameterized by analyzing the experimental data, i.e without using 
 nucleon PDFs or nuclear correction factor. This approach has been recently used by
 nCTEQ~\cite{Kovarik:2015cma, Kovarik:2012zz} group in getting $F_{2A}^{EM}(x,Q^2)$, $F_{2A}^{WI}(x,Q^2)$ 
 and $F_{3A}^{WI}(x,Q^2)$, who have collectively analyzed the charged lepton-$A$ DIS and DY $p-A$ dilepton production data sets~\cite{Kovarik:2015cma} to determine the 
 nuclear correction factor in the electromagnetic sector, and have performed an independent 
 analysis for the $\nu_l(\bar \nu_l)-A$ DIS data sets~\cite{Kovarik:2012zz}. It has been concluded by them
 that the nuclear medium effects in $F_{2A}^{EM}(x,Q^2)$ are different from $F_{2A}^{WI}(x,Q^2)$ 
 specially in the region of low $x$. Thus 
 in this region there is a disagreement 
 between the observation of these two studies~\cite{deFlorian:2011fp,Kovarik:2015cma}, specially at low $x$~\cite{Kalantarians:2017mkj}.
 
 Theoretically many models have been proposed to study these effects on the basis of nuclear binding, nuclear medium modification including short range
 correlations in nuclei~\cite{ Akulinichev1985, Dunne:1985ks, Bickerstaff:1989ch, CiofiDegliAtti:1989eg, Kulagin:1989mu, Arneodo:1992wf, Hen:2013oha, 
 Piller:1999wx, Marco:1995vb, Benhar:1997vy, Smith:2003hu, Kulagin:2004ie, CiofidegliAtti:2007vx, Kulagin:2007ju, SajjadAthar:2007bz, SajjadAthar:2009cr, Haider:2011qs,
 Frankfurt:2012qs, Haider:2012nf, Haider:2012ic,
 Haider:2014iia, Malace:2014uea, Ericson:1983um, Bickerstaff:1985mp, Berger1987, Jaffe:1982rr, Mineo:2003vc, Cloet:2005rt}, 
 pion excess in nuclei~\cite{Bickerstaff:1989ch, Kulagin:1989mu, Marco:1995vb, Ericson:1983um, Bickerstaff:1985mp, Berger1987}, 
 multi-quark clusters~\cite{Jaffe:1982rr, Mineo:2003vc, Cloet:2005rt}, 
 dynamical rescaling~\cite{Nachtmann:1983py, Close:1983tn}, nuclear shadowing~\cite{Frankfurt:1988nt, Armesto:2006ph}, etc.  Despite these efforts, no comprehensive theoretical/phenomenological
 understanding of the nuclear modifications of the  bound nucleon structure functions across the 
 complete range of $x$ and $Q^2$ consistent with the presently available experimental data exists~\cite{Arneodo:1992wf, Geesaman:1995yd, Hen:2013oha, Piller:1999wx}.
To understand nuclear modifications, theoretically various studies are available concerning the nuclear medium effects
 in the electromagnetic sector~\cite{Haider:2015vea, Zaidi:2019mfd, Hen:2013oha, Geesaman:1995yd} but there are mainly two groups, namely the group of Kulagin and 
 Petti~\cite{Kulagin:1989mu, Kulagin:2004ie, Kulagin:2007ju, Kulagin:2010gd} and Haider et al.~\cite{Haider:2016tev, Haider:2016zrk, Haider:2011qs, Haider:2012ic, Haider:2012nf} 
 who have made a comparative study of the nuclear medium effects in the electromagnetic and weak interaction induced processes~\cite{Haider:2016zrk}. 
 
As the nucleon structure functions are the basic inputs in the determination of nuclear structure functions and the scattering cross section, therefore, proper understanding of the nucleon structure functions as well as the parton
dynamics become quite important. In the region of low and moderate $Q^2$, the perturbative 
 and nonperturbative QCD corrections such as $Q^2$ evolution of parton distribution functions from leading order to higher order terms 
 (next-to-leading order (NLO), next-next-to-leading order (NNLO), ...), the effects of target mass correction due to 
 the massive quarks production (e.g. charm, bottom, top) and higher twist (twist-4, twist-6, ...) because of the multiparton correlations, become important. 
 These nonperturbative effects are specifically important in the kinematical region of high $x$ and low $Q^2$, sensitive to some of the oscillation parameters, and therefore it is of considerable
 experimental interest to the long baseline oscillation experiments.  
 
In this work, we have evaluated the nucleon structure functions by using the MMHT PDFs parameterization~\cite{Harland-Lang:2014zoa} up to next-to-next-to-leading order (NNLO) in the four
flavor($u,~d,~s,$ and $c$) scheme following Ref.~\cite{Vermaseren:2005qc, Moch:2004xu, Moch:2008fj}.
The nonperturbative higher twist
 effect is incorporated by using the renormalon approach~\cite{Dasgupta:1996hh} and the target mass correction is included following the works 
 of Schienbein et al.~\cite{Schienbein:2007gr}. 
After taking into account the QCD corrections at the free nucleon level, we have studied the modifications in the nuclear structure functions due to the presence of nuclear medium effects
such as Fermi motion, binding energy and nucleon correlations. These effects are incorporated through the use of spectral
function of the nucleon in the nuclear medium~\cite{Marco:1995vb, FernandezdeCordoba:1991wf}. The effect of mesonic contribution 
has been included which is found to be significant in the low and intermediate region of $x$~\cite{Marco:1995vb}. We have also included the effect of 
shadowing and antishadowing corrections following the works of Kulagin and Petti~\cite{Kulagin:2004ie}. Furthermore, we have discussed the effect of center 
of mass energy $(W)$ cut on $\nu_{l}-A$ and $\bar\nu_{l}-A$ scattering cross sections. 
 This paper is organized as follows.
 
 In the next section (section~\ref{sec_formalism}), we present the formalism in brief for (anti)neutrino-nucleon and 
  (anti)neutrino-nucleus DIS processes. Then we have discussed the method of obtaining nuclear structure functions with 
  medium effects such as Fermi motion, binding energy, nucleon correlations, mesonic contribution and shadowing.
 In section~\ref{sec_results}, numerical results are presented and discussed, and in the last section~\ref{summary} we summarize our findings.
\section{Formalism}\label{sec_formalism}
\subsection{Deep inelastic scattering of (anti)neutrino from nucleons}
The basic reaction for the (anti)neutrino induced charged current deep inelastic scattering process on a free nucleon target is given by
\begin{eqnarray}\label{reaction}
 \nu_l(k) / \bar\nu_l(k) + N(p) \rightarrow l^-(k') / l^+(k') + X(p')\;;\;\;\;l=e,\mu,
\end{eqnarray}
where $k$ and $k'$ are the four momenta of incoming and outgoing lepton, $p$ and $p'$ are the four momenta of the target nucleon and the jet of hadrons produced 
in the final state, respectively. This process is mediated by the $W$-boson ($W^\pm$) and the invariant matrix element corresponding to the above reaction is given by
\begin{equation}\label{matrix}
 -i{\cal M}=\frac{iG_F}{\sqrt{2}}\;l_\mu \;\left(\frac{M_W^2}{q^2-M_W^2} \right)\;\langle X|J^\mu|N\rangle\;.
\end{equation}
$G_F$ is Fermi coupling constant, $M_W$ is the mass of $W$ boson, and $q^2=(k-k')^2$ is the four momentum transfer square. $l_\mu$ is the leptonic current and $\langle X|J^\mu|N\rangle$ is the 
hadronic current for the neutrino induced reaction. The general expression of double differential scattering cross section (DCX) for the massless lepton limit ($m_l\to 0$) corresponding to the reaction given
in Eq.~\ref{reaction} in the laboratory frame is expressed as
\begin{equation}
\label{eq:w1w2w3}
\frac{ d^2\sigma_N^{WI} }{ dx dy } =  \frac{y M_N}{\pi }~\frac{E}{E'}~\frac{|{\bf k^\prime}|}{|{ \bf k}|}\;   \bar\sum \sum |{\cal M}|^2 \;,
\end{equation}
where $x=\frac{Q^2}{2 M_N \nu}$ is the Bjorken scaling variable, $y=\frac{p . q}{p.k}(=\frac{\nu}{E}~\rm{in~the~lab~frame})$ is the inelasticity, $\nu=E-E'$ is the energy transfer, $M_N$ is the nucleon mass, $E(E')$ is the energy of the incoming(outgoing) lepton and $\bar\sum \sum |{\cal M}|^2$ is the 
invariant matrix element square which is given in terms of the leptonic ($L_{\mu\nu}^{WI}$) and hadronic ($ W^{\mu\nu}_N$) tensors as
\begin{equation}\label{amp_wk}
 \bar\sum \sum |{\cal M}|^2 = \frac{G_F^2}{2}~\left(\frac{M_W^2}{Q^2+M_W^2}\right)^2 ~L_{\mu\nu}^{WI}~W^{\mu\nu}_N,
\end{equation}
with $Q^2=-q^2\ge 0$. $L_{\mu \nu}^{WI}$ is given by 
\begin{eqnarray}\label{lep_weak}
L_{\mu \nu}^{WI}&=&8(k_{\mu}k'_{\nu}+k_{\nu}k'_{\mu}
-k.k^\prime g_{\mu \nu}  \pm i \epsilon_{\mu \nu \rho \sigma} k^{\rho} 
k'^{\sigma})\,.
\end{eqnarray}
Here the antisymmetric term arises due to the contribution from the axial-vector components with +ve sign
for antineutrino and -ve sign for neutrino. 
\begin{figure}
\begin{center}
 \includegraphics[height=3.2 cm, width=11 cm]{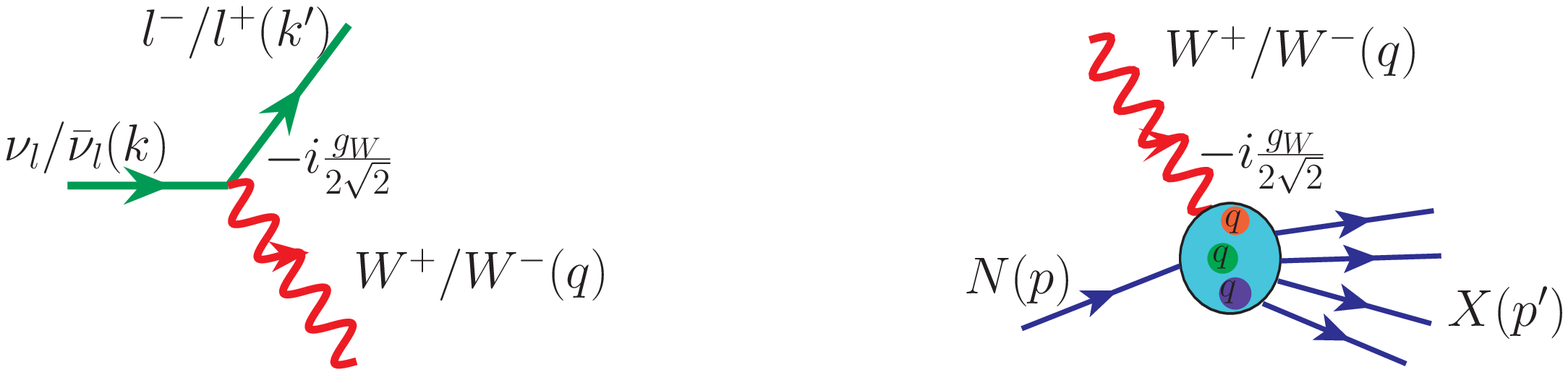}
  \end{center}
 \caption{Feynman representation for leptonic and hadronic vertices in the case of weak interaction.}
 \label{lag_weak}
\end{figure}
The hadronic tensor $W_{N}^{\mu \nu}$ is written in terms of the weak structure functions $W_{iN}^{WI}(\nu,Q^2)~(i=1-3)$ as
\begin{eqnarray}\label{had_weak_red}
W_{N}^{\mu \nu} 
&=&\left( \frac{q^{\mu} q^{\nu}}{q^2} - g^{\mu \nu} \right) \;
W_{1N}^{WI}(\nu,Q^2)
+ \frac{W_{2N}^{WI}(\nu,Q^2)}{M_N^2}\left( p^{\mu} - \frac{p . q}{q^2} \; q^{\mu} \right)
\left( p^{\nu} - \frac{p . q}{q^2} \; q^{\nu} \right)
 \nonumber\\
&&-\frac{i}{2M_N^2} \epsilon^{\mu \nu \rho \sigma} p_{ \rho} q_{\sigma}~
W_{3N}^{WI}(\nu,Q^2).
\end{eqnarray}
 The nucleon structure function $W_{3N}^{WI}(\nu,Q^2)$ arises due to the vector$-$axial vector interference part of 
the weak interaction and is responsible for the parity violation. 

The weak nucleon structure functions $W_{iN}^{WI} (\nu,Q^2)$(i=1,2,3) are generally redefined in terms of the dimensionless 
nucleon structure functions $F_{iN}^{WI}(x,Q^2)$ as: 
\begin{equation}\label{ch2:relation}
\left.
 \begin{array}{c}
M_N W_{1N}^{WI}(\nu, Q^2)=F_{1N}^{WI}(x,Q^2),\\
\nu W_{2N}^{WI}(\nu, Q^2)\;=\;F_{2N}^{WI}(x,Q^2),\\
\nu W_{3N}^{WI}(\nu, Q^2)\;=\;F_{3N}^{WI}(x,Q^2).
 \end{array}
 \right\}
\end{equation}
In general, the dimensionless nucleon structure functions are in turn written in terms of the parton distribution functions as
\begin{equation}\label{parton_wk}
\left.\begin{array}{c}
F_{2}^{WI}(x)  = \sum_{i} x [q_i(x) +\bar q_i(x)] \;,\\
x F_3^{WI}(x) =  \sum_i x [q_i(x) -\bar q_i(x)].
\end{array}
\right\}
\end{equation} 
In the above expressions, $i$ runs for the different flavors of quark(antiquark), the variable $x$ is the momentum fraction 
carried by a quark(antiquark) of the nucleon's momentum and $q_i(x)(\bar q_i(x))$ represents the probability density of finding a quark(antiquark) with a momentum fraction $x$. 
Using Eqs.~\ref{amp_wk}, \ref{lep_weak}, \ref{had_weak_red} and \ref{ch2:relation} in Eq.~\ref{eq:w1w2w3}, the differential scattering cross section is obtained as
\begin{eqnarray}\label{d2sigdxdy_weak}
 \frac{ d^2\sigma_N^{WI}}{ dx dy }&=& \frac{G_F^2 M_N E}{\pi} \left( \frac{M_W^2}{M_W^2+Q^2}\right)^2 \left[x y^2 F_{1N}^{WI}(x,Q^2) + \left(1-y-\frac{M_N x y}{ 2 E} \right) F_{2N}^{WI}(x,Q^2)\right.\nonumber\\
 &&  \left.  \pm x y\left(1-\frac{y}{2} \right)F_{3N}^{WI}(x,Q^2)\right]\;,
\end{eqnarray}
We have evaluated the nucleon structure functions up to NNLO following the works of Vermaseren et al.~\cite{Vermaseren:2005qc} and 
Moch et al.~\cite{Moch:2004xu, Moch:2008fj}. These structure functions
are expressed in terms of the convolution of coefficient function ($C_{a,f}\;;\;(f=q,g ~\textrm{and}~a=1-3)$) with the density distribution of partons ($f$) inside the nucleon. For example, we may write
$F_{2N}^{WI}(x)$ in terms of coefficient function as
\begin{equation}\label{f2_conv}
 x^{-1} F_{2N}^{WI}(x) = \sum_{f=q,g} C_{2,f}(x) \otimes f(x)\; ,
\end{equation}
with the perturbative expansion
\begin{equation}
 C_{2,f}(x)=\sum_{m}\left(\frac{\alpha_s(Q^2)}{2\pi}\right)^m\;c_{2,f}^{(m)},
\end{equation}
where superscript $m=0,1,2,...$ for N$^{(m)}$LO, $c_{2,f}^{(m)}(x)$ is the coefficient function for $F_{2N}^{WI}(x)$, $\alpha_s(Q^2)$ is the strong coupling constant and symbol $\otimes$ is the Mellin convolution which turns into simple multiplication in the N-space.
To obtain the convolution of coefficient functions with
parton density distribution, we use the following expression~\cite{Neerven}
\begin{equation}
 C_{a,f}(x)\otimes f(x) =\int_x^1 C_{a,f}(y)\; f\left(\frac{x}{y} \right) {dy \over y}
\end{equation}
The expression for the weak structure function $F_{3N}^{WI}(x)$ in terms of the coefficient function and parton density distribution function 
is given by~\cite{Moch:2008fj}: 
\begin{eqnarray}
 F_{3N}^{WI}(x) &=& \sum_{f=q,g} C_{3,f}(x) \otimes f(x)= C_{3,q}(x) \otimes q_v(x),\nonumber
\end{eqnarray}
where $q_v(x)(=f(x))$ is the valence quark distribution for a SU(3)/SU(4) symmetric sea and $C_{3,q}(x)$ is the coefficient function for $F_{3N}^{WI}(x)$. 

In the kinematic region of low and moderate $Q^2$, both the higher order perturbative and the nonperturbative ($\propto\frac{1}{Q^2}$) QCD effects come into play.
For example, the nonperturbative target mass correction effect involves the powers of $\frac{1}{Q^2}$, and is associated with the finite mass of the target nucleon. This effect is significant in the region of 
low $Q^2$ and high $x$ which is important to determine the valence quarks distribution. 
The higher twist (HT) effect which is suppressed by $\left(\frac{1}{Q^2}\right)^n~;~n=1,2,...$, originates due to the interactions of struck quarks with the other quarks present in the surroundings
via gluon exchange. This effect becomes small at low $x$ and high $Q^2$.
We have incorporated both the target mass correction and higher twist effects following Refs.~\cite{Schienbein:2007gr, Dasgupta:1996hh}
as well as performed the NNLO corrections in the evaluation of the nucleon structure functions.
For the numerical calculations, we have used the MMHT nucleonic PDFs parameterization~\cite{Harland-Lang:2014zoa}.
According to the operator product expansion~\cite{Wilson:1969zs, Shuryak:1981kj}, the weak nucleon structure functions with these nonperturbative effects can be mathematically expressed as 
\begin{equation}
 F_{iN}^{WI}(x,Q^2)=  F_{iN}^{WI,\tau=2}(x,Q^2) + \frac{H_{i}^{\tau=4}(x,Q^2)}{Q^2},
\end{equation}
where the leading twist term ($\tau=2$) incorporating the TMC effect obeys the Altarelli-Parisi evolution equations~\cite{Altarelli:1977zs}. It is written in terms of PDFs and is responsible for the evolution 
of structure functions via perturbative QCD $\alpha_s(Q^2)$ corrections. While the general expression of the twist-4 ($\tau=4$) term that reflects the strength of multi-parton 
correlations is given by~\cite{Dasgupta:1996hh}
\begin{eqnarray}
 H_{i}^{\tau=4}(x,Q^2)=A'_{2}\int_x^1 \frac{dz}{z}\;C^i_{2}(z)\;q\left(\frac{x}{z},Q^2\right),
\end{eqnarray}
with $i=1,2,3$. $C^i_{2}$ is the coefficient function for twist-4, $A'_{2}$ is the constant parameter and $q(x/z,Q^2)$ is the quarks density distribution.

We have incorporated the medium effects using a microscopic field theoretical approach. The effect of Fermi motion, binding energy and nucleon correlations 
 are included through the relativistic nucleon spectral function which is 
 obtained by using the Lehmann's representation for the relativistic nucleon propagator.  We use the technique of nuclear many body theory to calculate the dressed nucleon propagator
 in an interacting Fermi sea in the nuclear matter. To obtain the results for a finite nucleus the local density approximation (LDA) is then applied. In the LDA, Fermi momentum of an interacting
 nucleon is not a constant 
 quantity but the function of position coordinate ($r$)~\cite{FernandezdeCordoba:1991wf}. Since the nucleons bound inside a nucleus interact among themselves via the exchange of virtual mesons 
 such as $\pi,~\rho,$ etc., therefore a finite probability of the interaction of intermediate vector boson with these mesons exists. We have also incorporated the mesonic contribution by using many-body field theoretical approach similar to the 
 case of bound nucleons~\cite{Marco:1995vb}. Furthermore, the shadowing effect is taken into account that dominates in the region of low $x$, where 
 the hadronization of intermediate vector bosons ($W^+/W^-$) creates quark-antiquark pairs that interact with the partons.
 The multiple scattering of quarks causes the destructive interference of amplitudes that leads to the phenomenon of 
 shadowing which is incorporated in this paper, following the works of Kulagin and Petti~\cite{Kulagin:2004ie}. In the next subsection,
 we have discussed the formalism adopted for the (anti)neutrino-nucleus scattering process.
\subsection{Deep inelastic scattering of (anti)neutrino from nuclei}
In the case of DIS of (anti)neutrino from nuclear targets the expression of the differential cross section is given by
\begin{eqnarray}\label{d2sig_wi}
\frac{ d^2\sigma^{WI}_A}{ dx dy }&=& \frac{G_F^2 \; M_N\; y}{ 2\pi}\;\frac{E}{E'} \frac{|{\bf k^\prime}|}{|{ \bf k}|} \left(\frac{M_W^2}{ M_W^2+Q^2 }\right)^2 L_{\mu\nu}^{WI}\; W^{\mu\nu}_A\;,
\end{eqnarray}
where $L_{\mu\nu}^{WI}$ is the weak leptonic tensor which has the same form as given in Eq.~\ref{lep_weak} while the nuclear hadronic tensor $W^{\mu\nu}_A$ 
is written in terms of the weak nuclear structure functions $W_{iA}^{WI}(\nu,Q^2)$ ($i=1,2,3$) relevant in the case of $m_l\to 0$ as
\begin{eqnarray}
 \label{nuc_had_weak}
 W^{\mu\nu}_A &=& \left( \frac{q^{\mu} q^{\nu}}{q^2} - g^{\mu \nu} \right) \;
W_{1A}^{WI}(\nu,Q^2)
+ \frac{W_{2A}^{WI}(\nu,Q^2)}{M_A^2}\left( p_A^{\mu} - \frac{p_A . q}{q^2} \; q^{\mu} \right)
\left( p_A^{\nu} - \frac{p_A . q}{q^2} \; q^{\nu} \right)
 \nonumber\\
&&-\frac{i}{2M_A^2} \epsilon^{\mu \nu \rho \sigma} p_{A \rho} q_{\sigma}~
W_{3A}^{WI}(\nu,Q^2).
\end{eqnarray}
After contracting the leptonic tensor with the hadronic tensor and using the following relations between the nuclear structure 
functions ($W_{iA}^{WI}(\nu,Q^2)$) and the dimensionless nuclear structure functions ($F_{iA}^{WI}(x,Q^2)$)
\begin{eqnarray} 
\label{eq:f1w1nua}
 M_A ~W_{1A}^{WI}(\nu,Q^2) &=& F_{1A}^{WI}(x,Q^2) \ , \\
\label{eq:f2w2nua}
  \nu ~W_{2A}^{WI}(\nu,Q^2) &=& F_{2A}^{WI}(x,Q^2) \ , \\
\label{eq:f3w3nua}
  \nu ~W_{3A}^{WI}(\nu,Q^2) &=& F_{3A}^{WI}(x,Q^2) \ ,
\end{eqnarray}
we obtain
\begin{eqnarray}\label{d2sigdxdy_weak1}
\frac { d^2\sigma_A^{WI} }{ dx dy }&=& \frac{G_F^2 M_N E}{\pi} \left( \frac{M_W^2}{M_W^2+Q^2}\right)^2 \left[x y^2 F_{1A}^{WI}(x,Q^2) + \left(1-y-\frac{M_N x y}{ 2 E} \right) F_{2A}^{WI}(x,Q^2)\right.\nonumber\\
 &&  \left.  \pm x y \left(1-\frac{y}{2} \right)F_{3A}^{WI}(x,Q^2)\right]\;.
\end{eqnarray}

When the interaction takes place with a nucleon bound inside a nucleus, it gets influenced by the presence of other nucleons which are not stationary but are continuously moving with a finite Fermi momentum.
This motion of nucleons corresponds to the Fermi motion. These bound nucleons may also interact among themselves via strong interaction that is incorporated by the nucleon-nucleon correlations and the 
binding energy for a given nucleus has also been ensured. Moreover, for a nonsymmetric nucleus such as iron, copper, tin, lead, etc., we have taken into account the different densities for the 
proton and the neutron. We have discussed these effects and present the formalism in the following subsection.
 \subsubsection{Fermi motion, binding energy, nucleon correlation and isoscalarity effects}
 To calculate the scattering cross section for a neutrino interacting 
with a target nucleon in the nuclear medium, we express it in terms of the probability of interaction per unit area which is defined as the probability of interaction per unit time
of the particle ($\Gamma$) times
the time spent in the interaction process ($dt$) over a differential area $dS$~\cite{Zaidi:2019mfd, Haider:2016zrk, Haider:2015vea}, i.e.
    \begin{eqnarray}\label{defxsec}
d\sigma&=&\Gamma \;dt\; dS=\Gamma \frac{1}{v}\;d^3 r=\Gamma \;\frac {E({\bf k})}{\mid {\bf k} \mid} \;d^3 r,
\end{eqnarray}
where $v\left(=\frac {\mid {\bf k} \mid} {E({\bf k})}\right)$ is the velocity of the particle and $d^3 r$ is the volume element.
The probability of interaction per unit time($\Gamma$) that the incoming neutrino will interact with the bound nucleons is
 related to the neutrino self-energy, which provides information about the total neutrino flux available 
at our disposal after the interaction:
  \begin{equation}\label{pro1}
 \Gamma=-\frac{2 m_l}{E({\bf k})}Im\Sigma~ \;\;\;\Rightarrow \;\;d\sigma=\frac{-2 m_l}{|{\bf k}|} Im\Sigma~d^3 r~,
\end{equation}
  where $Im \Sigma$ stands for the imaginary part of the neutrino self-energy that accounts for the depletion of the initial neutrinos flux out of the non-interacting 
  channel, into the quasielastic or the inelastic channels.
 Thus the imaginary part of the neutrino self-energy gives information about the total number of neutrinos that have participated in the interaction and give rise to the charged 
 leptons. 
Therefore, the evaluation of imaginary part of the neutrino self-energy is required to obtain the scattering cross section. Following the Feynman rules 
we write the neutrino self-energy corresponding to the diagram shown in Fig.~\ref{wself_energy}(a) as:
\begin{figure}
\begin{center}
 \includegraphics[height=4.0 cm, width=10 cm]{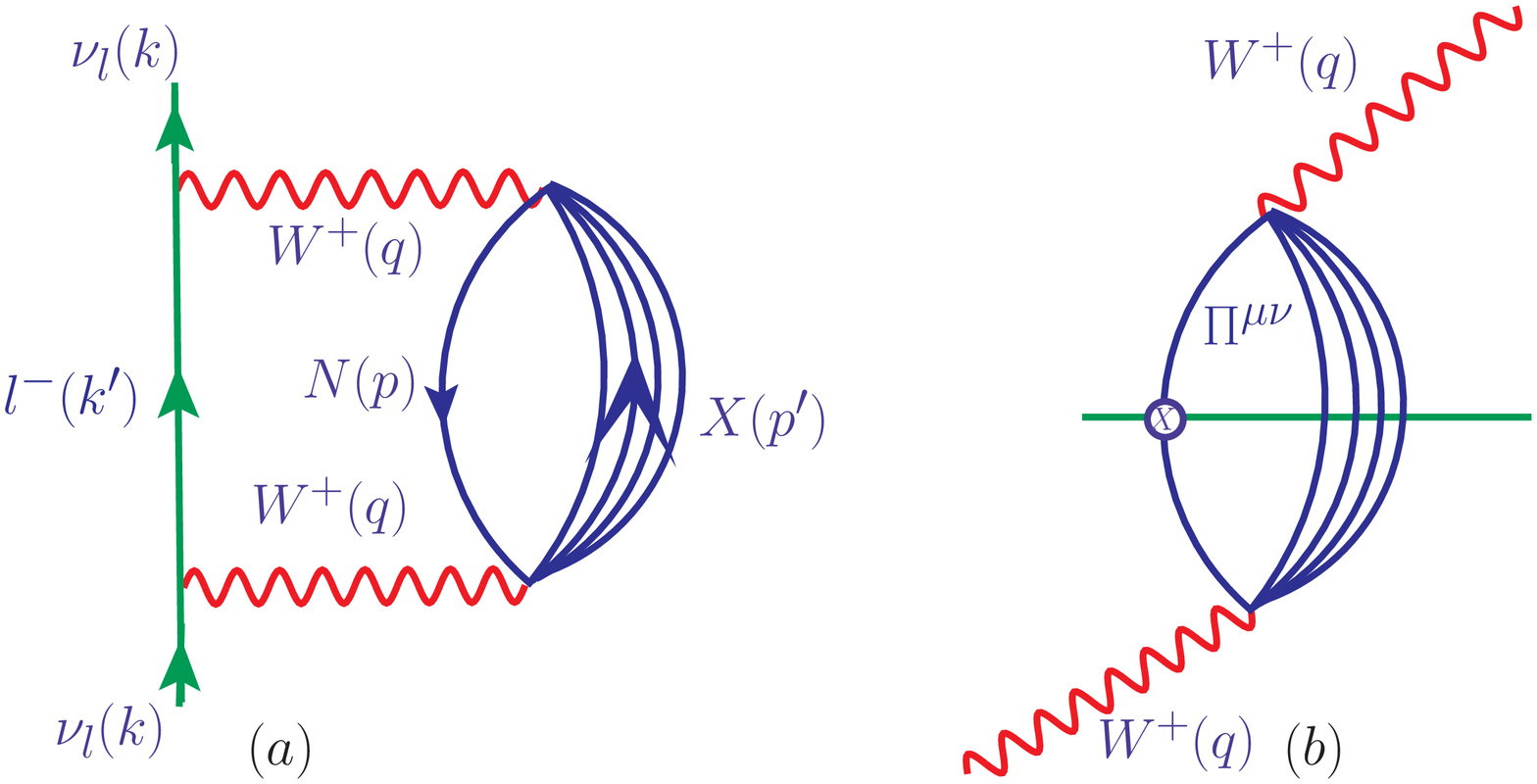}
 \end{center}
 \caption{Diagrammatic representation of the neutrino self-energy (left panel) and intermediate $W$ boson self-energy (right panel).}
 \label{wself_energy}
\end{figure}
\begin{eqnarray}
\Sigma(k)&=&\frac{- i G_F}{\sqrt{2}} \int \frac{d^4 k^\prime}{(2 \pi)^4} \frac{4 L_{\mu\nu}^{WI}}{m_l}\frac{1}{(k^{\prime 2}-m_l^2+i\epsilon)}\left(\frac{M_W}{Q^2+M_W^2}\right)^2 \Pi^{\mu\nu}(q)\;,
\end{eqnarray}
where we have used the properties of gamma matrices.
Imaginary part of the neutrino self-energy may be obtained by using the Cutkosky rules~\citep{Haider:2016zrk} and is given by
\begin{equation}\label{nu_imslf}
Im \Sigma(k)=\frac{ G_F}{\sqrt{2}} \frac{4}{ m_l} \int \frac{d^4 k^\prime}{(2 \pi)^4} \frac{\pi}{ E'({\bf k^\prime})} \theta(q^0) L_{\mu\nu}^{WI }\left(\frac{M_W}{Q^2+M_W^2}\right)^2\;Im[\Pi^{\mu\nu}(q)].
\end{equation}
In the above expression, $\Pi^{\mu\nu}(q)$ is the $W$ boson self-energy (depicted in Fig.~\ref{wself_energy}(b)) which is defined in terms of the intermediate nucleon ($G_l$) and meson ($D_j$) propagators:
\begin{eqnarray}\label{wboson}
 \Pi^{\mu \nu} (q)&=& \left(\frac{G_F M_W^2}{\sqrt{2}}\right) \times \int \frac{d^4 p}{(2 \pi)^4} G (p) 
\sum_X \; \sum_{s_p, s_l} \prod_{i = 1}^{N} \int \frac{d^4 p'_i}{(2 \pi)^4} \; \prod_{_l} G_l (p'_l)\; \prod_{_j} \; D_j (p'_j)\nonumber \\  
&&  <X | J^{\mu} | N >  <X | J^{\nu} | N >^* (2 \pi)^4  \; \delta^4 ( p +q - \sum^N_{i = 1} p'_i),\;\;\;
\end{eqnarray}
where $s_p$ is the spin of the nucleon, $s_l$ is the spin of the fermions in $X$, $<X | J^{\mu} | N >$ is the hadronic current for the initial state nucleon 
to the final state hadrons, index $l,~j$ are respectively, stands for the fermions and the bosons in the final hadronic state $X$ and $\delta^4 (p +q  - \sum^N_{i = 1} p'_i)$ ensures the conservation
of four momentum at the vertex.
$G(p)$ is the nucleon propagator inside the nuclear medium through which the information about the propagation of the nucleon from the initial state to the final state or vice versa is obtained. 
The relativistic nucleon propagator for a noninteracting Fermi sea is written in terms of the positive ($u(p)$) and negative ($v(-p)$) energy components as:
\begin{eqnarray}  \label{prop4}
G^{0}(p^{0},{{\bf p}})&=&\frac{M_N}{E_N({{\bf p}})}\left\{\sum_{r}u_{r}({\bf p})\bar u_{r}({\bf p})
\left[\frac{1-n(\bf{p})}{p^{0}-E_N({{\bf p}})+i\epsilon}+\frac{n(\bf{p})}{p^{0}-E_N({{\bf p}})-i\epsilon}\right]
 +\frac{\sum_{r}v_{r}(-{\bf p})\bar v_{r}(-{\bf p})}{p^{0}+E_N({{\bf p}})-i\epsilon}\right\}.\nonumber
\end{eqnarray}
The nucleon propagator retains the contribution only from the positive energy components because the negative energy components are much suppressed. Hence, we obtain
\begin{eqnarray}  \label{prop45}
G^{0}(p^{0},{{\bf p}})&=&\frac{M_N}{E_N({{\bf p}})}\sum_{r}u_{r}({\bf p})\bar u_{r}({\bf p})
\left[\frac{1-n(\bf{p})}{p^{0}-E_N({{\bf p}})+i\epsilon}+\frac{n(\bf{p})}{p^{0}-E_N({{\bf p}})-i\epsilon}\right]. \nonumber
\end{eqnarray}
In the above expression, the first term of the nucleon propagator within the square bracket contributes when the momentum of nucleon will be greater or equal to the Fermi momentum ${\bf |p|} \ge {\bf p_{F}}$, i.e. for the particles above the Fermi sea
while the second term within the square bracket contributes when the nucleon momentum will be less than the Fermi momentum ${\bf |p|} < {\bf p_{F}}$, i.e. for the particles below the Fermi
sea. This representation is known as the Lehmann's representation\cite{Marco:1995vb}. 
\begin{figure}
\begin{center}
\includegraphics[width=10cm,height=4cm]{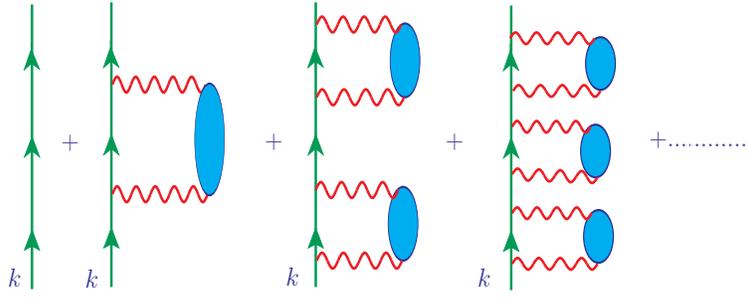}
\end{center}
\caption{Diagrammatic representation of nucleon self-energy in the nuclear medium.}
\label{nuc_self}
\end{figure}
Inside the Fermi sea, where nucleons interact with each other, the relativistic nucleon propagator
$G(p)$ is obtained by using the perturbative expansion of Dyson series in terms of the nucleon self energy$(\Sigma^N)$ as:
\begin{eqnarray}\label{gpseries}
 G(p) &=& G^0(p)~+~G^0(p)\Sigma^N(p)G^0(p)~+~G^0(p)\Sigma^N(p)G^0(p)\Sigma^N(p)G^0(p)~+~...\;\;.\nonumber
\end{eqnarray}
The nucleon self energy (shown in Fig.\ref{nuc_self}) is evaluated by using the many body field theoretical approach in terms of the
spectral functions~\cite{Marco:1995vb, FernandezdeCordoba:1991wf} and the dressed nucleon propagator $G(p)$ in an interacting Fermi sea is obtained as~\cite{FernandezdeCordoba:1991wf}: 
\begin{eqnarray}\label{Gp}
G (p) = \frac{M_N}{E_N({\bf p})} 
\sum_r u_r ({\bf p}) \bar{u}_r({\bf p})
\left[\int^{\mu}_{- \infty} d  \omega 
\frac{S_h (\omega, {\bf{p}})}{p^0 - \omega - i \epsilon}
+ \int^{\infty}_{\mu} d  \omega 
\frac{S_p (\omega, {\bf{p}})}{p^0 - \omega + i \epsilon}\right]\,,
\end{eqnarray}
where $\mu=\epsilon_F+M_N$ is the chemical potential,
$\omega=p^0-M_N$ is the removal energy, $S_h (\omega, {\bf{p}})$ and $S_p (\omega, {\bf{p}})$ are the hole
and particle spectral functions, respectively. In the above expression the term $S_h (\omega, {\bf{p}})~d\omega$ is basically the joint probability of 
removing a nucleon from the ground state and $S_p (\omega, {\bf{p}})~d\omega$ is the joint probability of adding a nucleon to the ground state of a nucleus. 
Consequently, one may obtain the spectral function sum rule which is given by
\begin{equation}
 \int_{-\infty}^{\mu}~S_h (\omega, {\bf{p}})~d\omega~+~ \int_{\mu}^{+\infty}~S_p (\omega, {\bf{p}})~d\omega~=~1.
\end{equation}
The expressions for the hole and particle spectral functions are given by~\cite{Marco:1995vb, FernandezdeCordoba:1991wf}:
\begin{equation}\label{sh}
 S_h(p^0,\mathbf{p})=\frac{1}{\pi}
 \frac{\frac{M_N}{E_N(\mathbf{p})}\textrm{Im}\Sigma^N(p^0,\mathbf{p})}{\left(p^0-
 E_N(\mathbf{p})-\frac{M_N}{E_N(\mathbf{p})}\textrm{Re}\Sigma^N(p^0,\mathbf{p})\right)^2+
 \left(\frac{M_N}{E_N(\mathbf{p})}\textrm{Im}\Sigma^N(p^0,\mathbf{p})\right)^2}
\end{equation}
when $p^0 \le \mu$,
\begin{equation}\label{sp}
 S_p(p^0,\mathbf{p})=-\frac{1}{\pi}
 \frac{\frac{M_N}{E_N(\mathbf{p})}\textrm{Im}\Sigma^N(p^0,\mathbf{p})}{\left(p^0-
 E_N(\mathbf{p})-\frac{M_N}{E_N(\mathbf{p})}\textrm{Re}\Sigma^N(p^0,\mathbf{p})\right)^2+
 \left(\frac{M_N}{E_N(\mathbf{p})}\textrm{Im}\Sigma^N(p^0,\mathbf{p})\right)^2}
\end{equation}
when $p^0 > \mu$. 

In the present study, we are considering the inclusive DIS process and are not looking at the final hadronic state, therefore, the interactions in the Fermi sea are taken into account through
the hole spectral function $S_h$. Now by using
Eqs.~\ref{pro1} and \ref{nu_imslf}, and performing the momentum space integration the differential scattering cross section is obtained as: 
\begin{equation}\label{dsigma_3}
\frac {d\sigma_A^{WI}}{dx dy}=-\frac{G_F^2\;M_N\;y}{2\pi}\;\frac{E}{E'}\;\frac{|\bf{k^\prime}|}{|\bf {k}|}\left(\frac{M_W^2}{Q^2+M_W^2}\right)^2 L_{\mu\nu}^{WI} \int  Im \Pi^{\mu\nu}(q) d^{3}r.
\end{equation}
On comparing Eq.~\ref{d2sig_wi} and Eq.~\ref{dsigma_3}, it is found that the nuclear hadronic tensor $W_A^{\mu \nu}$ is related with the imaginary part of the $W$ boson self-energy
$Im \Pi^{\mu\nu}(q)$ as
\begin{equation}\label{wamunu}
W_A^{\mu \nu}=- \int  Im \Pi^{\mu\nu}(q) d^{3}r.
\end{equation}
 Using Eq.~\ref{Gp} and the expressions for the nucleon and meson propagators in Eq.~\ref{wboson}, and finally substituting 
 them in Eq.~\ref{wamunu}, we obtain the nuclear hadronic tensor $W^{\mu \nu}_{A}$ for an isospin symmetric nucleus in terms of the nucleonic hadronic tensor $W^{\mu \nu}_{N}$
 convoluted with the hole spectral function($S_h$) for a nucleon bound inside the nucleus:
\begin{equation}\label{conv_WA}
W^{\mu \nu}_{A} = 4 \int \, d^3 r \, \int \frac{d^3 p}{(2 \pi)^3} \, 
\frac{M_N}{E_N ({\bf p})} \, \int^{\mu}_{- \infty} d p^0 S_h (p^0, {\bf p}, \rho(r))
W^{\mu \nu}_{N} (p, q)~,
\end{equation}
where the factor of 4 is for spin-isospin of the nucleon and $\rho(r)$ is the nuclear density. In general, nuclear density have various phenomenological parameterizations known 
in the literature as the harmonic oscillator(HO) density, two parameter Fermi density(2pF),
modified harmonic oscillator (MHO) density, etc. The proton density distributions are obtained from the electron-nucleus scattering experiments, while the neutron densities
are taken from the Hartee-Fock approach~\cite{GarciaRecio:1991wk}. The density parameters $c_1$ and $c_2$ corresponds to the charge density for proton or equivalently the neutron matter density for neutron. In the present model, for the numerical calculations, we have used modified harmonic oscillator charge density 
 \begin{equation}\label{mho}
  \rho(r)=\rho_0 \left[1 + c_2 \left(\frac{r}{c_1} \right)^2\right] e^{-\left(\frac{r}{c_1}\right)^2 } 
 \end{equation}
 for the light nuclei, e.g. $^{12}$C, and 2-parameter Fermi density
 \begin{equation}\label{2para}
  \rho(r)=\frac{\rho_0} {\left[1+e^{\left(\frac{r-c_1}{c_2}\right)}\right]}
 \end{equation}
 for the heavy nuclei, like $^{40}$Ar, $^{56}$Fe and $^{208}$Pb. In Eqs.~\ref{mho} and \ref{2para}, $\rho_0$ is the central 
 density and $c_1$, $c_2$ are the density parameters~\cite{DeJager:1987qc, GarciaRecio:1991wk} which are independently given for protons ($c_{1,2}^p$) and neutrons ($c_{1,2}^n$) in Table~\ref{tab:nuc_para} along with the 
 other parameters used in the numerical calculations. 
  \begin{table}
 \centering
\begin{tabular}{c|cc|cc|c|c|c} \hline
 \multirow{2}{*}{{ Nucleus}}& \multicolumn{2}{c|}{ $ {c_1}$ }  &   \multicolumn{2}{c|}{ $ {c_2}$ } &  \multirow{2}{*}{$<r^2>^{1/2}$} & $ \multirow{2}{*}{ {B.E./A}}$  & $ \multirow{2}{*}{ {T/A}}$\\\cline{2-5}
   &    $ {c_1^n}$ & $ {c_1^p}$ & $ {c_2^n}$ & $ {c_2^p}$  &  &   &   \\\hline  
{$^{12}$C}  &  1.692   & 1.692    & 1.075$^{\ast}$   &   1.075$^{\ast}$     &  2.47 & 7.5   &  26.0 \\
{$^{40}$Ar} &  3.53   & 3.53    &  0.542     &   0.542        & 3.393 & 8.6    & 29.0  \\
{$^{56}$Fe} &  4.050  & 3.971   &  0.5935    &   0.5935     &  3.721 & 8.8    &  30.0   \\
 {$^{208}$Pb} &  6.890   & 6.624    &  0.549   &  0.549     &  5.521 & 7.8    & 32.6 \\ \hline 
\end{tabular}
\caption{Different parameters used for the numerical calculations for various nuclei. For $^{12}$C
we have used modified harmonic oscillator density($^{\ast}$ $c_2$ is dimensionless) and for $^{40}$Ar, $^{56}$Fe and $^{208}$Pb nuclei,
 2-parameter Fermi density have been used, where superscript $n$ and $p$ in density parameters($c_{i}^{n,p}$; $i$=1,2) stand for neutron and proton, respectively. Density parameters and the 
 root mean square radius ($<r^2>^{1/2}$) are given in units of femtometer. The kinetic energy of the nucleon per nucleus($T/A$) and binding energy of the nucleon per nucleus ($B.E/A$)
 for different nuclei are given in MeV. }
 \label{tab:nuc_para}
\end{table}
We ensure the normalization of the hole spectral function by obtaining the baryon number ($A$) of a given nucleus
and binding energy of the same nucleus. 
\begin{eqnarray}
  4 \int d^3r\;\int \frac{d^3 p}{(2\pi)^3} \;\int_{-\infty}^{\mu}\;S_h(\omega,{\bf p},\rho(r))\;d\omega &=& A\;, \nonumber
 \end{eqnarray}
In the local density approximation, the spectral functions for the proton ($Z$) and neutron ($N=A-Z$) numbers in a nuclear target which are the function of local Fermi momenta
 $p_{_{{F}_{p,n}}}(r)=\left[ 3\pi^{2} \rho_{p(n)}({r}) \right]^{1/3}$, are normalized separately such that
\begin{eqnarray}
  2 \int d^3r\;\int \frac{d^3 p}{(2\pi)^3} \;\int_{-\infty}^{\mu_p}\;S_h^p(\omega,{\bf p},\rho_p(r))\;d\omega &=& Z\;, \nonumber\\
    2 \int d^3r\;\int \frac{d^3 p}{(2\pi)^3} \;\int_{-\infty}^{\mu_n}\;S_h^n(\omega,{\bf p},\rho_n(r))\;d\omega &=& N\;, \nonumber
 \end{eqnarray}
where the factor of 2 is due to the two possible projections of nucleon spin, $\mu_p(\mu_n)$ is the chemical potential for the proton(neutron), and $S_h^p(\omega,{\bf p},\rho_p(r))$ and
$S_h^n(\omega,{\bf p},\rho_n(r))$ are the hole spectral functions for the proton and neutron, respectively.
 The proton and neutron densities $\rho_{p}(r)$ and $\rho_{n}(r)$ are related to the nuclear density $\rho(r)$ as~\cite{Haider:2015vea, Haider:2016zrk}: 
\begin{eqnarray}
 \rho_{p}(r) &=& \frac{Z}{A}\;\rho(r)~;\hspace{5 mm}
 \rho_{n}(r) = \frac{(A-Z)}{A}\;\rho(r)\nonumber
\end{eqnarray}
Hence for a nonisoscalar nuclear target, the nuclear hadronic tensor is written as
\begin{equation}\label{conv_WAan}
W^{\mu \nu}_{A} = 2 \sum_{\tau=p,n} \int \, d^3 r \, \int \frac{d^3 p}{(2 \pi)^3} \, 
\frac{M_N}{E_N ({\bf p})} \, \int^{\mu_\tau}_{- \infty} d p^0 S_h^\tau (p^0, {\bf p}, \rho_\tau(r))\;
W^{\mu \nu}_{N} (p, q). \,
\end{equation}
In this way, we have incorporated the effects of Fermi motion, Pauli blocking and nucleon correlations through the hole spectral function.

From Eqs.~\ref{conv_WA} and \ref{conv_WAan}, we have evaluated the nuclear structure functions by using the expressions of nucleon and nuclear
hadronic tensors given in Eqs.~\ref{had_weak_red} and \ref{nuc_had_weak}, respectively with the suitable choice of their components along $x,y,$ and $z$ axes. 
The numerical calculations are performed in the laboratory frame, where the target nucleus is assumed to be at rest($p_A$=($p_A^0$,${\bf p_A}=0$)) but the nucleons are moving with finite
momentum(p=($p^0$,${\bf p}\ne 0$)). These nucleons are thus off shell. If we choose the momentum transfer (${\bf q}$) to be along the $z$ axis, i.e, $q^\mu=(q^0,0,0,q^z)$.
Then the Bjorken variables for the nuclear target and the bound nucleons are defined as
\begin{eqnarray}
x_A&=&\frac{Q^2}{2 p_A \cdot q}= \frac{Q^2}{2 M_{A}  q^0}= \frac{Q^2}{2 A~M_N q^0}\;, \hspace{6 mm}
x_N = \frac{Q^2}{2 p \cdot q} = \frac{Q^2}{2 (p^0 q^0 - p^z q^z)}.
\end{eqnarray}

Hence, we have obtained the expressions of weak nuclear structure functions for the isoscalar and nonisoscalar nuclear targets by using Eqs.~\ref{conv_WA} and \ref{conv_WAan}, respectively.
The expression of $F_{1A,N}^{WI}(x_A, Q^2)$ is obtained by taking the $xx$ component of nucleon (Eq.~\ref{had_weak_red}) and nuclear (Eq.~\ref{nuc_had_weak}) hadronic tensors which
for an isoscalar nuclear target is given by
\begin{eqnarray}
\label{conv_WA1_wk_iso}
F_{1A,N}^{WI}(x_A, Q^2) &=& 4AM_N \int \, d^3 r \, \int \frac{d^3 p}{(2 \pi)^3} \, 
\frac{M_N}{E_N ({\bf p})} \, \int^{\mu}_{- \infty} d p^0~ S_h(p^0, {\bf p}, \rho(r))~\nonumber\\
&\times& \left[\frac{F_{1N}^{WI}(x_N, Q^2)}{M_N} + \left(\frac{p^x}{M_N}\right)^2 \frac{F_{2N}^{WI}(x_N, Q^2)}{\nu_N}\right],\\
&&\nonumber
\end{eqnarray}   
and for a nonisoscalar nuclear target is obtained as
\begin{eqnarray}	\label{conv_WA2weak}
F_{1A,N}^{WI}(x_A, Q^2) &=& 2\sum_{\tau=p,n} AM_N \int \, d^3 r \, \int \frac{d^3 p}{(2 \pi)^3} \, 
\frac{M_N}{E_N ({\bf p})} \, \int^{\mu_\tau}_{- \infty} d p^0 ~S_h^\tau (p^0, {\bf p}, \rho_\tau(r))~ \nonumber\\
&\times&\left[\frac{F_{1\tau}^{WI}(x_N, Q^2)}{M_N} + \left(\frac{p^x}{M_N}\right)^2 \frac{F_{2\tau}^{WI}(x_N, Q^2)}{\nu_N}\right],
 \end{eqnarray}
 where $\nu_N=\frac{p\cdot q}{M_N}=\frac{p^0 q^0 - p^z q^z}{M_N}$.
We must point out that the evaluation of $F_{1A,N}^{WI}(x_A, Q^2)$ has been performed independently, i.e., without using the Callan-Gross relation at the nuclear level.
Similarly, the $zz$ component of nucleon (Eq.~\ref{had_weak_red}) and nuclear (Eq.~\ref{nuc_had_weak}) hadronic tensors gives the expression of 
dimensionless nuclear structure function $F_{2A,N}^{WI}(x_A,Q^2)$. For an isoscalar nuclear target it is expressed as
 \begin{eqnarray}
  \label{had_ten151_wk_iso}
F_{2A,N}^{WI}(x_A,Q^2)  &=&  4 \int \, d^3 r \, \int \frac{d^3 p}{(2 \pi)^3} \, 
\frac{M_N}{E_N ({\bf p})} \, \int^{\mu}_{- \infty} d p^0 ~S_h (p^0, {\bf p}, \rho(r)) \nonumber \\
&\times& \left[\frac{Q^2}{(q^z)^2}\left( \frac{|{\bf p}|^2~-~(p^{z})^2}{2M_N^2}\right) +  \frac{(p^0~-~p^z~\gamma)^2}{M_N^2} \left(\frac{p^z~Q^2}{(p^0~-~p^z~\gamma) q^0 q^z}~+~1\right)^2\right]~\nonumber\\
&\times&\left(\frac{M_N}{p^0~-~p^z~\gamma}\right) \times F_{2 N}^{WI}(x_N,Q^2),       
\end{eqnarray}
 while for a nonisoscalar nuclear target it modifies to
  \begin{eqnarray} 
\label{had_ten151weak}
F_{2A,N}^{WI}(x_A,Q^2)  &=&  2\sum_{\tau=p,n} \int \, d^3 r \, \int \frac{d^3 p}{(2 \pi)^3} \, 
\frac{M_N}{E_N ({\bf p})} \, \int^{\mu_\tau}_{- \infty} d p^0 ~S_h^\tau (p^0, {\bf p}, \rho_\tau(r)) \nonumber\\
&\times&\left[\left(\frac{Q}{q^z}\right)^2\left( \frac{|{\bf p}|^2~-~(p^{z})^2}{2M_N^2}\right) +  \frac{(p^0~-~p^z~\gamma)^2}{M_N^2} \left(\frac{p^z~Q^2}{(p^0~-~p^z~\gamma) q^0 q^z}~+~1\right)^2\right]  \nonumber \\
&\times&\left(\frac{M_N}{p^0~-~p^z~\gamma}\right) \times F_{2\tau}^{WI}(x_N,Q^2),   
\end{eqnarray}
with $\gamma=\frac{q^0}{q^z}$.

The expression of $F_{3A,N}^{WI}(x_A,Q^2)$ is obtained by choosing the $xy$ component of nucleon (Eq.~\ref{had_weak_red}) and nuclear (Eq.~\ref{nuc_had_weak}) hadronic tensors
which is given by
\begin{eqnarray}\label{f3a_weak}
 F_{3A,N}^{WI}(x_A,Q^2) &=& 4 A  \int \, d^3 r \, \int \frac{d^3 p}{(2 \pi)^3} \, 
\frac{M_N}{E_N ({\bf p})} \, \int^{\mu}_{- \infty} d p^0 S_h (p^0, {\bf p}, \rho(r))\times \frac{q^0}{q^z} \nonumber\\
&\times& \left({p^0 q^z - p^z q^0  \over p \cdot q} \right)F_{3N}^{WI}(x_N,Q^2),
\end{eqnarray}
for an isoscalar nuclear target. However, for a nonisoscalar nuclear target, we get
\begin{eqnarray}\label{f3a_weak_noniso}
 F_{3A,N}^{WI}(x_A,Q^2) &=& 2 A \sum_{\tau=p,n} \int \, d^3 r \, \int \frac{d^3 p}{(2 \pi)^3} \, 
\frac{M_N}{E_N ({\bf p})} \, \int^{\mu_\tau}_{- \infty} d p^0 S_h^\tau (p^0, {\bf p}, \rho_\tau(r))\times \frac{q^0}{q^z} \nonumber\\
&\times& \left({p^0 q^z - p^z q^0  \over p \cdot q} \right)F_{3\tau}^{WI}(x_N,Q^2),
\end{eqnarray}
The results obtained by using Eqs.~\ref{conv_WA1_wk_iso}, \ref{had_ten151_wk_iso}, \ref{f3a_weak} for isoscalar and Eqs.~\ref{conv_WA2weak}, \ref{had_ten151weak}, \ref{f3a_weak_noniso} for nonisoscalar nuclear targets 
are labeled as the results with the spectral function(SF) only. 
\subsubsection{Mesonic Effect}
\begin{figure}
\begin{center} 
 \includegraphics[height=0.2\textheight,width=0.88\textwidth]{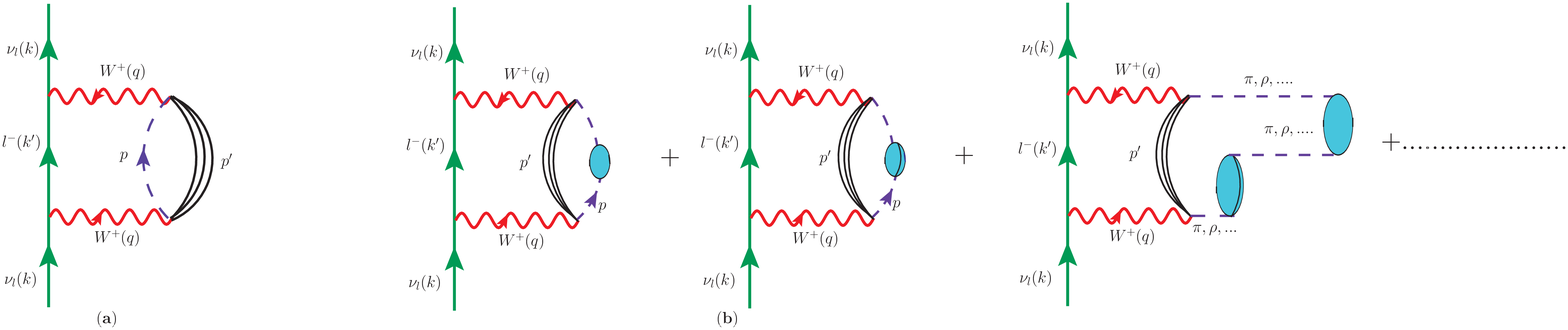}
 \caption{Neutrino self-energy diagram accounting for neutrino-meson DIS (a) the bound nucleon propagator is substituted with a meson($\pi$ or $\rho$) propagator of momentum $p$ and 
 jet of hadrons $X$ with momentum $p'$.
 (b) by including particle-hole $(1p–1h)$, delta-hole $(1\Delta–1h)$, $1p1h-1\Delta1h$, etc. interactions.}
 \label{meson}
 \end{center}
\end{figure}

In the case of (anti)neutrino-nucleus DIS process, mesonic effects also contribute to the nuclear structure functions $F_{1A}^{WI}(x_A, Q^2)$ and $F_{2A}^{WI}(x,Q^2)$ 
which arises due to the interaction of bound nucleons among themselves via the exchange of virtual mesons such as $\pi,~\rho,$ etc. 
There is a reasonably good probability that intermediate $W$ boson may interact with a meson instead of a nucleon~\cite{Marco:1995vb,Kulagin:2004ie}.
In order to include the contribution from the virtual mesons, we again evaluate the neutrino self-energy for which a diagram is shown in Fig.~\ref{meson} and write the 
meson hadronic tensor in the nuclear medium similar to the case of bound nucleons as \cite{Marco:1995vb}
\begin{equation}\label{W2pion}
W^{\mu \nu}_{A, i} = 3 \int d^3 r \; \int \frac{d^4 p}{(2 \pi)^4} \;
\theta (p_0) (- 2) \; Im D_i (p) \; 2 m_i W^{\mu \nu}_{i} (p, q)\; ,
\end{equation}
where $i=\pi, \rho$, a factor of 3 is due to the three charged states of pion (rho meson) and $D_i(p)$ is the dressed meson propagator. This expression is obtained by replacing the hole spectral function
\[-2\pi\frac{M_N}{E_N(\mathbf{p})}S_{h}(p_0,\mathbf{p})~W^{\mu \nu}_{N} (p, q)\] in 
 Eq.~\ref{conv_WA} with the imaginary part of the meson propagator, i.e, \[ImD_i(p)\; \theta(p_0)\; 2 W^{\mu \nu}_{i} (p, q).\]
 This meson propagator does not correspond to the free mesons because a lepton (either electron or muon) can not decay into another lepton, one pion and debris of hadrons
 but it corresponds to the mesons arising due to the nuclear medium 
 effects by using a modified meson propagator. These mesons are arising in the nuclear medium through 
 particle-hole(1p-1h), delta-hole(1$\Delta$-1h), 1p1h-1$\Delta$1h, 2p-2h, etc. interactions as depicted in Fig.~\ref{meson}.
This effect is incorporated following the mean-field theoretical approach~\cite{Marco:1995vb}. 
 The expression of meson propagator ($D_i(p)$) in the nuclear medium is given by 
 \begin{equation}\label{dpi}
D_i(p) = [ {p_0}^2 - {\bf {p}}\,^{2} - m^2_{i} - \Pi_{i} (p_0, {\bf p}) ]^{- 1}\,,
\end{equation}
with the mass of meson $m_i$ and the meson self-energy $\Pi_i$ which is explicitly written as
\begin{equation}\label{pionSelfenergy}
\Pi_\pi=\left({f^2 \over m_\pi^2}\right)\;\frac{ F^2_\pi(p){\bf {p}}\,^{2}\Pi^*(p)}{1-\left({f^2 \over m_\pi^2}\right) V'_L(p) \Pi^*(p)}\,,\hspace{5 mm} \Pi_\rho=\left({f^2 \over m_\pi^2}\right)\;\frac{C_\rho F^2_\rho(p){\bf {p}}\,^{2}\Pi^*(p)}{1-\left({f^2 \over m_\pi^2}\right) V'_T(p) \Pi^*(p)}\,.
\end{equation}
In the above expressions, the coupling constant $f=1.01$, the free parameter $C_\rho=3.94$, $V'_L(p) (V'_T(p))$ is
the longitudinal(transverse) part of the spin-isospin interaction which is responsible for the enhancement to the pion(rho meson) structure function and $\Pi^*(p)$ is the irreducible meson
self-energy that contains the contribution of particle-hole and delta-hole excitations.
The $\pi NN$ and $\rho NN$ form factors, i.e., $F_\pi(p)$ and $F_\rho(p)$ used in Eq.~\ref{pionSelfenergy} are given by
\begin{equation}\label{fpi}
 F_\pi(p)={(\Lambda_\pi^2-m_\pi^2)\over (\Lambda_\pi^2+{\bf {p}}\,^{2})}\;,\hspace{6 mm}  F_\rho(p)={(\Lambda_\rho^2-m_\rho^2)\over (\Lambda_\rho^2+{\bf {p}}\,^{2})}
\end{equation}
with the parameter $\Lambda_\pi(\Lambda_\rho)$=1~GeV. 
Since Eq.\ref{W2pion} has taken into account the mesonic contents of the nucleon which are already incorporated in the sea contribution of the nucleon, in order to calculate the 
mesonic excess in the nuclear medium we have subtracted the meson contribution of the nucleon~\cite{Marco:1995vb} such that
\begin{equation}
Im D_i (p) \; \rightarrow \; \delta I m D_i (p) \equiv I m D_i (p) - \rho \;
\left.\frac{\partial Im D_i (p)}{\partial \rho} \right|_{\rho = 0} .
\end{equation}
Now we have obtained the following expression for the mesonic hadronic tensor
\begin{equation}\label{W2piona}
W^{\mu \nu}_{A, i} = 3 \int d^3 r \; \int \frac{d^4 p}{(2 \pi)^4} \;
\theta (p_0) (- 2) \; \delta Im D_i (p) \; 2 m_i W^{\mu \nu}_{i} (p, q)\; ,
\end{equation}
Using Eq.\ref{W2piona}, the mesonic structure functions $F_{1A, i}^{WI}(x,Q^2)$ and $F_{2A, i}^{WI}(x,Q^2)$ are evaluated following 
the same analogy as adopted in the case of bound nucleons~\cite{Marco:1995vb}. The expression for $F_{1A, i}^{WI}(x,Q^2)$ is given by
\begin{eqnarray} 
\label{F2rho_wk}
F_{1A, i}^{WI}(x,Q^2) &=& - 6\times a \times A M_N \int  d^3 r   \int  \frac{d^4 p}{(2 \pi)^4} \theta (p^0) ~\delta I m D_i (p) \;2m_i\nonumber\\
&\times& ~\left[\frac{F_{1i}^{WI}(x_i)}{m_i}~+~\frac{{|{\bf p}|^2~-~(p^{z})^2}}{2(p^0~q^0~-~p^z~q^z)}
\frac{F_{2i}^{WI}(x_i)}{m_i}\right],
\end{eqnarray}
and for $F_{2A, i}^{WI}(x,Q^2)$ we obtain
\begin{eqnarray} 
 \label{F2rho1_wk}
F_{2A, i}^{WI}(x,Q^2)  &=& - 6\times a \int \, d^3 r \, \int \frac{d^4 p}{(2 \pi)^4} \, 
        \theta (p^0) ~\delta I m D_i (p) \;2m_i\nonumber \\
&\times&\left[\frac{Q^2}{(q^z)^2}\left( \frac{|{\bf p}|^2~-~(p^{z})^2}{2m_i^2}\right)  
+  \frac{(p^0~-~p^z~\gamma)^2}{m_i^2} \left(\frac{p^z~Q^2}{(p^0~-~p^z~\gamma) q^0 q^z}~+~1\right)^2\right]~\nonumber\\
&\times& \left(\frac{m_i}{p^0~-~p^z~\gamma}\right) ~F_{2i}^{WI}(x_i),
\end{eqnarray}
where $x_i=\frac{Q^2}{-2p \cdot q}$ and $a=1$ for pion and $a=2$ for rho meson~\cite{Marco:1995vb}. Notice that the $\rho$ meson has an extra factor of two compared to pionic 
contribution because of the two transverse polarization of the $\rho$ meson~\cite{Baym:1975vb}.

In the literature, various groups like MRST98~\cite{Martin:1998sq}, CTEQ5L~\cite{Wijesooriya:2005ir}, SMRS~\cite{Sutton:1991ay}, GRV\cite{Gluck:1991ey}, etc., have proposed the 
quark and antiquark PDFs parameterizations for pions. We have observed in our earlier work~\cite{Zaidi:2019mfd} that the choice of different pionic PDFs parameterization would not 
make much difference in the scattering cross section. For the present numerical 
calculations the GRV pionic PDFs parameterization given by Gluck et al.~\cite{Gluck:1991ey} has been used and 
the same PDFs are also taken for the rho meson. The contribution from the pion cloud is found to be larger than the contribution from rho meson cloud, nevertheless, the
rho contribution is non-negligible, and both of them are positive in the whole range of $x$. 
It is important to mention that $F_{3A}^{WI}(x_A, Q^2)$ has no 
mesonic contribution as it depends mainly on the valence quarks distribution and these average to zero when considering the three charge states of pions and rho mesons. For details
please see Refs.~\cite{Zaidi:2019mfd, Haider:2015vea, Haider:2016zrk}.
\subsubsection{Shadowing and Antishadowing effects}
The shadowing effect which contributes in the region of low $x(\le 0.1)$, takes place as a result of the destructive interference of the amplitudes due to the multiple scattering of quarks
arising due to the hadronization of $W^\pm/Z^0$ bosons
and leads to a reduction in the nuclear structure functions.
It arises when the coherence length is larger than the average distance
between the nucleons bound inside the nucleus and the expected coherence time is $\tau_c\ge  2$ fm. However, the shadowing effect gets saturated if the coherence length becomes larger
than the average nuclear radius, i.e., in the region of low $x$. 
Furthermore, in the region of $0.1<x<0.3$, the nuclear structure functions get enhanced due to the antishadowing effect
which is theoretically less understood. In the literature, several studies 
proposed that it may be associated with the constructive interference of scattering amplitudes resulting from the multiple scattering of quarks~\cite{Kulagin:2007ju, Kulagin:2004ie, Kopeliovich:2012kw}. For the antishadowing effect, the coherence
time is small for the long inter-nucleon spacing in the nucleus corresponding to these values of $x$. 
Shadowing and antishadowing effects are found to be quantitatively different in electromagnetic and weak interaction induced processes~\cite{Haider:2016zrk}. It is because the electromagnetic and weak interactions
take place through the interaction of photons and $W^\pm/Z^0$ bosons, respectively, with the target hadrons and 
 the hadronization processes of photons and $W^\pm/Z^0$ bosons are different. Moreover, in the case of weak interaction, 
 the additional contribution of axial current which is not present in the case of electromagnetic interaction may 
 influence the behaviour of weak nuclear structure functions specially if pions also play a role in the hadronization process through PCAC.
 Furthermore, in this region of low $x$, sea quarks also play an important role which could be different in the case of electromagnetic and
weak processes. 
In the present numerical calculations, we have incorporated the shadowing effect following the works 
of Kulagin and Petti~\cite{Kulagin:2004ie} who have used Glauber-Gribov multiple scattering theory. For example, to determine the nuclear structure function $F_{iA}^{WI}(x,Q^2)$
with the shadowing effect, we use~\cite{Kulagin:2004ie}
\begin{equation}
 F_{iA}^{WI,S}(x,Q^2) = \delta R(x,Q^2) \times F_{iN}^{WI}(x,Q^2)\; ,
 \label{shdw11}
\end{equation}
where $F_{iA}^{WI,S}(x,Q^2);~(i=1-3)$ is the nuclear structure function with shadowing effect and the factor $\delta R(x,Q^2)$ is given in Ref.~\cite{Kulagin:2004ie}.

Now, using the present formalism, we have presented the results for the weak structure functions and scattering 
cross sections for both the free nucleon and nuclear targets in the next section. 
\section{Results and Discussion}\label{sec_results}
We have performed the numerical calculations by considering the following cases:
\begin{itemize}
\item The nucleon structure functions are obtained using PDFs parameterization of Martin et al.~\cite{Harland-Lang:2014zoa}.
\item All the results are presented with TMC effect.
 \item $F_{iN}^{WI}(x,Q^2);~(i=1-3)$ are obtained at NLO and NNLO.
 \item At NLO the higher twist effect has been incorporated following the renormalon approach~\cite{Dasgupta:1996hh} and a
 comparison is made with the results obtained at NNLO.
 \item After taking into account the perturbative and nonperturbative QCD corrections in the evaluation of free nucleon structure functions, we have used them to calculate the 
 nuclear structure functions. The expression for $F_{iA}^{WI}(x,Q^2),~(i=1,2)$ in the full model is given by
\begin{equation}\label{sf_full}
  F_{iA}^{WI}(x,Q^2)= F_{iA,N}^{WI}(x,Q^2) + \;F_{iA, \pi}^{WI}(x,Q^2)  + F_{iA, \rho}^{WI}(x,Q^2) \;+ F_{iA}^{WI,S}(x,Q^2)\;,
\end{equation}
where $F_{iA,N}^{WI}(x,Q^2)$ is the structure function with spectral function given in
Eqs.\ref{conv_WA1_wk_iso}(\ref{conv_WA2weak}) and \ref{had_ten151_wk_iso}(\ref{had_ten151weak}) for $F_{1A,N}^{WI}(x,Q^2)$ and $F_{2A,N}^{WI}(x,Q^2)$, respectively,
in the case of isoscalar (nonisoscalar)
targets which takes care of Fermi motion, binding energy and nucleon correlations. The mesonic contributions are included using Eq.\ref{F2rho_wk} and \ref{F2rho1_wk} for $F_{1A,j}^{WI}(x,Q^2)$
and $F_{2A,j}^{WI}(x,Q^2)$ ($j=\pi,\rho$), respectively and for the shadowing effect ($F_{iA}^{WI,S}(x,Q^2)$) Eq.\ref{shdw11} is used. 
$F_{3A}^{WI}(x,Q^2)$ has no mesonic contribution and the expression is given by 
\begin{eqnarray}\label{f3_tot}
 F_{3A}^{WI}(x,Q^2)= F_{3A,N}^{WI}(x,Q^2) + F_{3A,shd}^{WI}(x,Q^2),
\end{eqnarray}
with spectral function contribution $F_{3A,N}^{WI}(x,Q^2)$ using Eqs.\ref{f3a_weak}(\ref{f3a_weak_noniso}) for the isoscalar (nonisoscalar) nuclear
targets and the shadowing correction $F_{3A,shd}^{WI}(x,Q^2)$ using Eq.\ref{shdw11}.
\begin{figure}
\begin{center}
 \includegraphics[height= 7.8 cm , width= 0.9\textwidth]{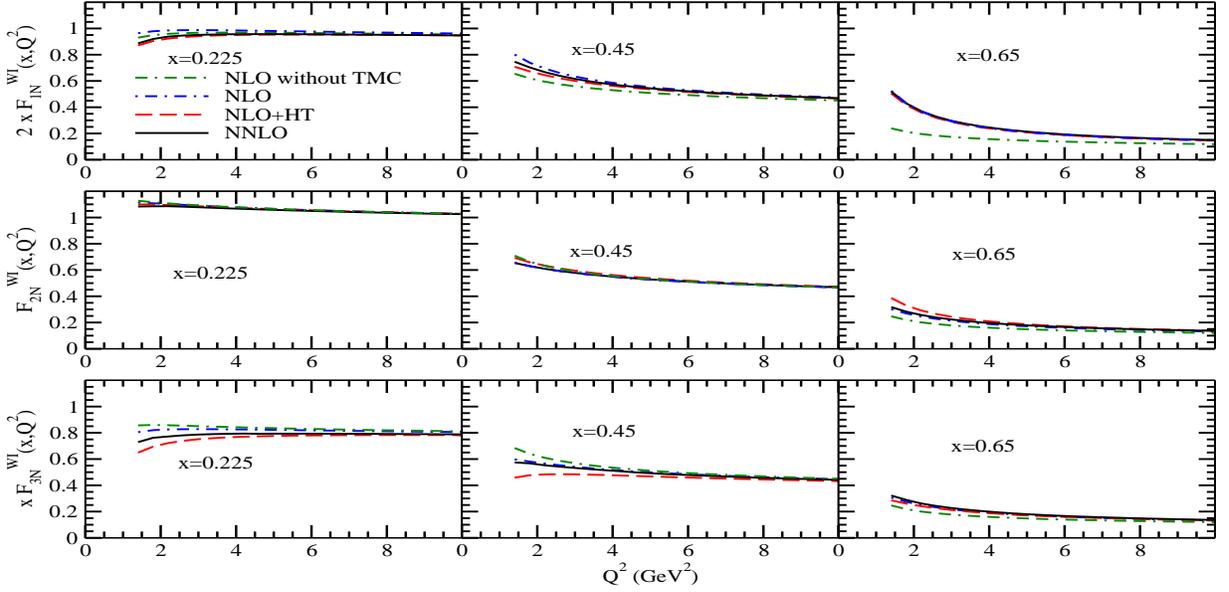}
\end{center}
\caption{$2 x F_{1N}^{WI}(x,Q^2)$ (top panel), $F_{2N}^{WI}(x,Q^2)$ (middle panel) and $F_{3N}^{WI}(x,Q^2)$ (bottom panel) vs $Q^2$ at the different values of $x$ 
for $\nu_l - N$ scattering. The results are obtained at NLO {\bf (i)} without TMC effect (double dash-dotted line), {\bf (ii)} including TMC effect
 without (dash-dotted line) and with (dashed line) the higher twist correction, and {\bf (iii)} at NNLO with TMC effect (solid line).}
\label{fig0}
\end{figure}
 \item The results are presented for $^{12}$C, CH, $^{40}$Ar, $^{56}$Fe and $^{208}$Pb nuclear targets which are being used in the present generation experiments.
\end{itemize}

The results of the free nucleon structure functions are presented in Fig.\ref{fig0}, for $2 x F_{1N}^{WI}(x,Q^2)$, 
$F_{2N}^{WI}(x,Q^2)$ and $F_{3N}^{WI}(x,Q^2)$ vs $Q^2$ at $x=0.225,0.45$ and $0.65$ in the case of neutrino-nucleon DIS process. We observe that due to the TMC effect the nucleon 
structure functions are modified at low and moderate $Q^2$ specially in the region of high $x$. We find that at NLO, the modification in the structure functions due to TMC effect
is about
$3\%(16\%)$ in $2 x F_{1N}^{WI}(x,Q^2)$, $<1\%(5\%)$ in $F_{2N}^{WI}(x,Q^2)$ and $5\%(10\%)$ in $F_{3N}^{WI}(x,Q^2)$
at $x=0.225(0.45)$ and $Q^2=1.8$ GeV$^2$ which becomes $1\%(8\%)$, $<1\%(1\%)$ and $\sim 2\%(3\%)$ at $Q^2=5$ GeV$^2$. 
On the other hand the effect of higher twist corrections in this kinematic region is very small in $F_{1N}^{WI}(x,Q^2)$ and $F_{2N}^{WI}(x,Q^2)$ unlike in the case of 
electromagnetic structure functions~\cite{Zaidi:2019mfd}. Whereas the effect of higher twist in $F_{3N}^{WI}(x,Q^2)$
leads to a decrease of $15\%$ at $x=0.225$ and $5\%$ at $x=0.65$ 
for $Q^2=1.8$ GeV$^2$, and becomes small with the increase in $Q^2$.  
We observe that the difference in the results of $F_{iN}^{WI}(x,Q^2)~(i=1,2)$ at NLO with HT effect from the results at NNLO is $<1\%$. However,
in $F_{3N}^{WI}(x,Q^2)$ at $x=0.225$, this difference is about $8\%$ for $Q^2=1.8$ GeV$^2$ and 
it reduces to $\sim 2\%$ for $Q^2=5$ GeV$^2$. With the increase in $x$ and $Q^2$ the effect becomes gradually smaller.
\begin{figure}
\begin{center}
 \includegraphics[height= 7.8 cm , width= 0.95\textwidth]{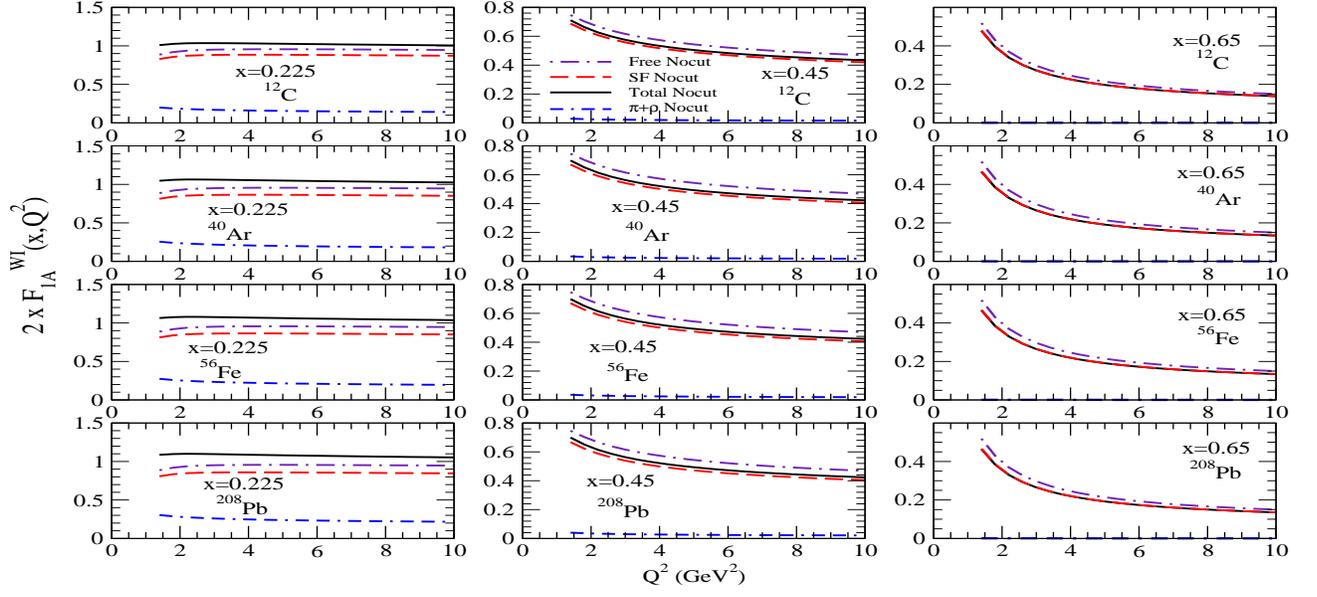}
\end{center}
\caption{Results for $2 x F_{1A}^{WI}(x,Q^2)$ vs $Q^2$ are shown at different values of $x$ for neutrino induced DIS process in $^{12}$C, $^{40}$Ar, $^{56}$Fe
and $^{208}$Pb. The results are obtained with the spectral function only (dashed line), with mesonic contribution only (double dash-dotted line)
and with the full model (solid line) at NNLO and are compared with the results of the
free nucleon case (dash-dotted line). All the nuclear targets are treated as isoscalar.}
\label{fig1}
\end{figure}

The effect of higher twist is further suppressed in the nuclear medium, which is similar to our observation made for the electromagnetic nuclear structure functions~\cite{Zaidi:2019mfd}. The 
results observed at NLO with higher twist is close to the results obtained at NNLO.
Therefore, all the results are presented here at NNLO. 
\begin{figure}
\begin{center}
 \includegraphics[height= 8 cm , width= 0.95\textwidth]{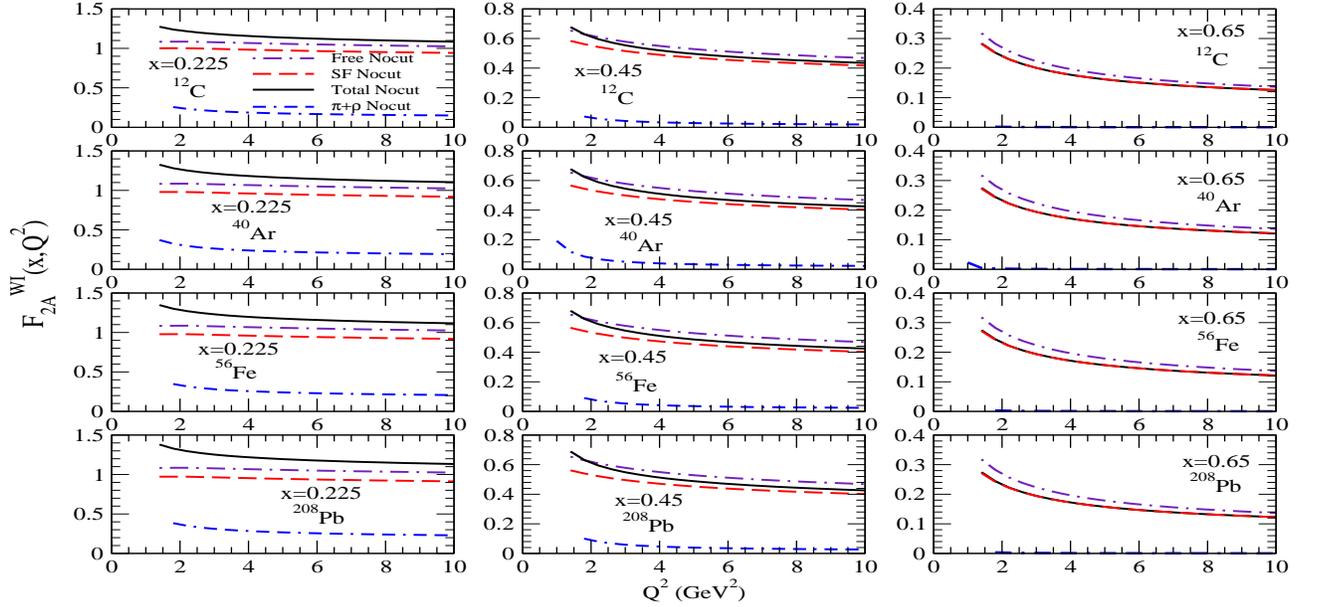}
\end{center}
\caption{$ F_{2A}^{WI}(x,Q^2)$ vs $Q^2$. The lines in this figure have the same meaning as in Fig.\ref{fig1}.}
\label{fig2}
\end{figure}
\begin{figure}
\begin{center}
 \includegraphics[height= 8 cm , width= 0.95\textwidth]{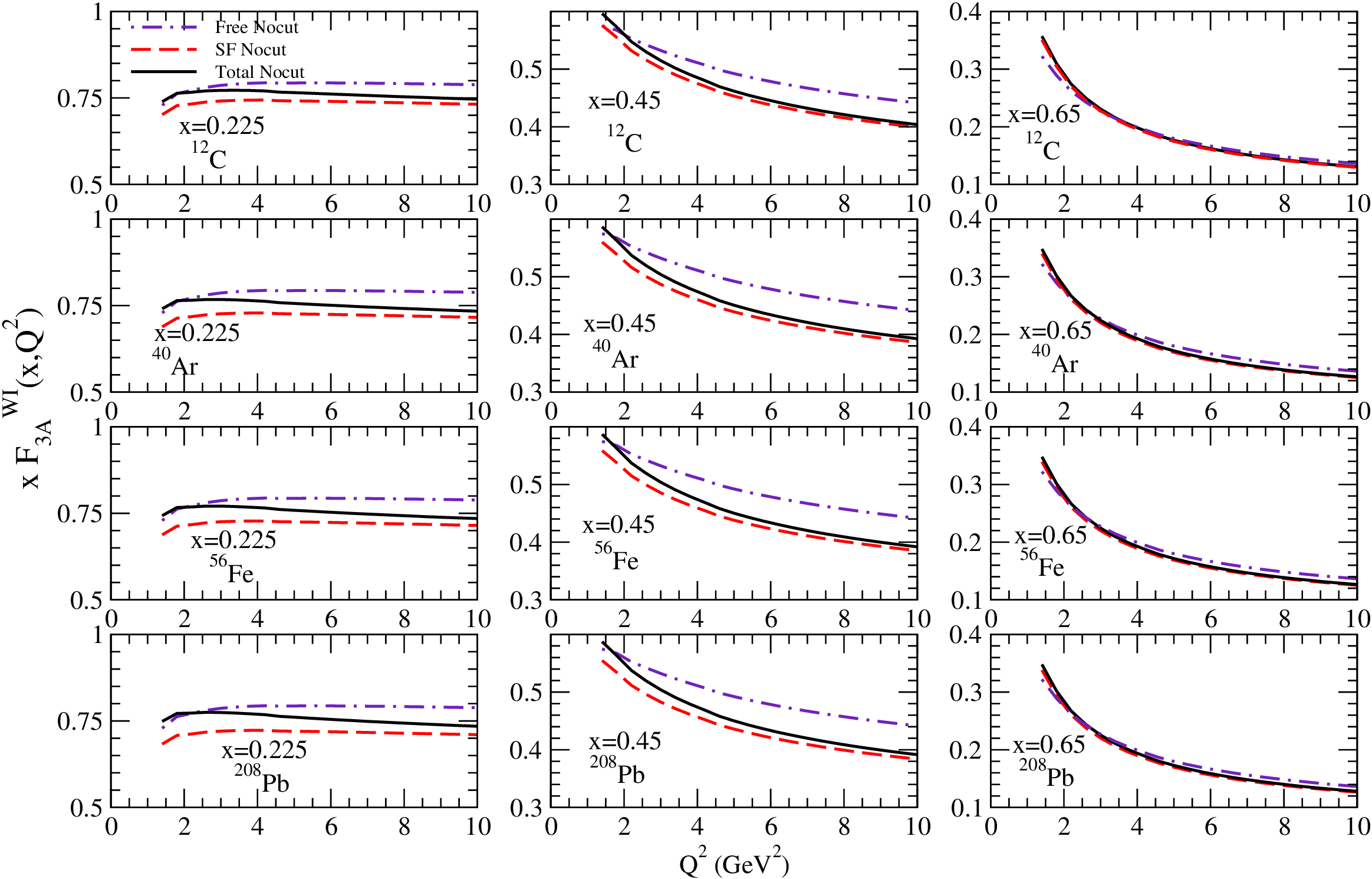}
\end{center}
\caption{ $ x F_{3A}^{WI}(x,Q^2)$ vs $Q^2$. The lines in this figure have the same meaning as in Fig.\ref{fig1}.}
\label{fig3}
\end{figure}

In Figs.~\ref{fig1}, \ref{fig2}, and \ref{fig3} the results are presented respectively for the nuclear structure functions $2 x F_{1A}^{WI}(x,Q^2)$, $ F_{2A}^{WI}(x,Q^2)$ and $x F_{3A}^{WI}(x,Q^2)$
vs $Q^2$ for different values of $x$.  The numerical results obtained in the kinematic limit $Q^2>1$ GeV$^2$ without any cut on the center of mass energy $W$ are
labeled as ``Nocut''.
The nuclear structure functions are shown for $1 <Q^2 \le 10$ GeV$^2$ in carbon, argon, iron and lead which are treated as isoscalar nuclear targets
and compared these results from the results obtained for a free nucleon target. From the figures, the different behaviour of the nuclear medium effects in different regions of $x$ and $Q^2$ can be clearly observed. For example, 
the results for the structure functions with spectral function are suppressed from the results of the free nucleon target in the range of $x(<0.7)$ and $Q^2$ considered here. 
Quantitatively, this reduction in carbon from the results of free nucleon structure functions
for $Q^2=1.8$ GeV$^2$ is found to be about $ 7\%$, $8\%$, and $\sim 5\%$ at $x=0.225$ in 
$2 x F_{1A}^{WI}(x,Q^2)$, $ F_{2A}^{WI}(x,Q^2)$, and $x F_{3A}^{WI}(x,Q^2)$ respectively, which becomes $9\%$, $11\%$, and $2\%$ at $x=0.45$. 
We have explicitly shown the mesonic contribution (double dash-dotted line) which is quite significant in the low and intermediate regions of $x(<0.6)$. The inclusion of mesonic effect 
gives an enhancement in the case of nuclear structure functions $F_{1A}^{WI}(x,Q^2)$ and $F_{2A}^{WI}(x,Q^2)$ for all values of $x<0.6$ and becomes negligible for $x> 0.6$. 
 The shadowing (antishadowing) effect that causes a reduction(enhancement) in the nuclear structure function for $x\le 0.1(0.1<x<0.3)$ is modulated by 
the mesonic contribution that works in its opposite (same) direction
and results in an overall enhancement of the nuclear structure functions.
Hence, the results obtained by including mesonic contributions, shadowing and antishadowing effects in our full model are higher than the 
results with the spectral function only. Mesonic contribution does not contribute to $x F_{3A}^{WI}(x,Q^2)$. 
The difference between the results of spectral function and the full model for $2 x F_{1A}^{WI}(x,Q^2)$ is $20\%$ at $x=0.225$ and $3\%$ at $x=0.45$ for $Q^2=1.8$ GeV$^2$ in carbon. 
These nuclear effects are observed to be more pronounced for the heavy nuclear targets such as in the case of argon 
it becomes $26\%$($4\%$) and $31\% (5\%)$ in lead at $x=0.225$($x=0.45$) for $Q^2=1.8$ GeV$^2$. However, with the increase in $Q^2$ the mesonic contribution becomes 
small, for example, at $Q^2=5$ GeV$^2$ this difference is reduced to $16\%$ in $^{12}C$, $21\%$ in $^{40}Ar$ and $26\%$ in $^{208}Pb$ at $x=0.225$. 

\begin{figure}
\begin{center}
 \includegraphics[height= 7.5 cm , width= 0.95\textwidth]{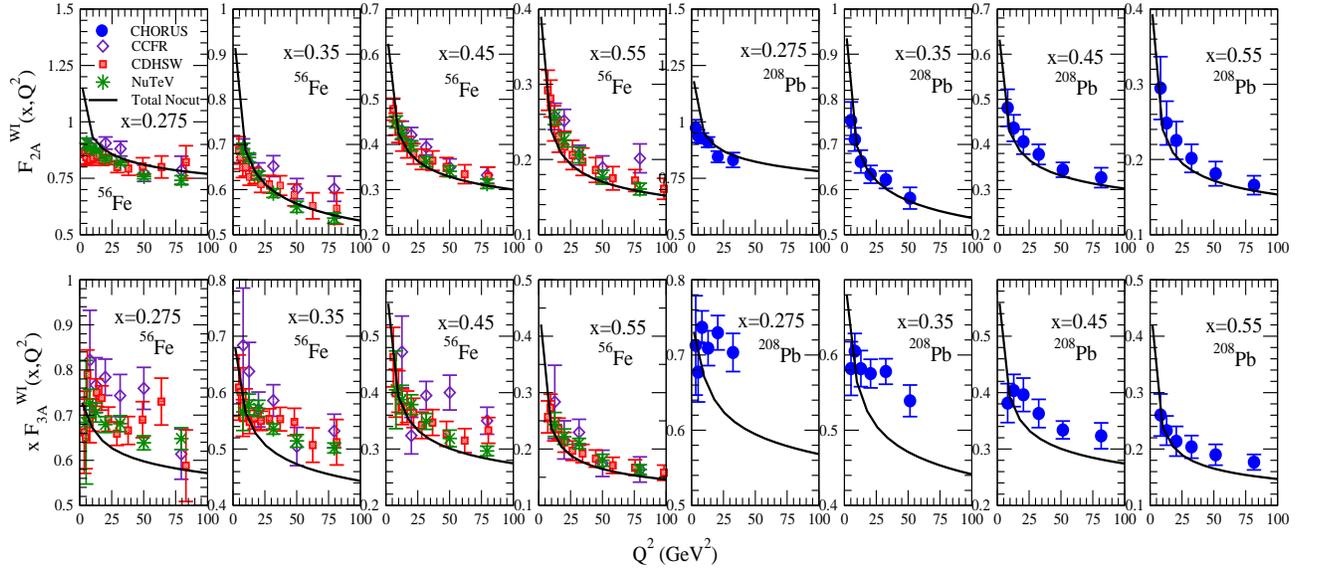}
\end{center}
\caption{Results for $ F_{2A}^{\nu+\bar\nu}(x,Q^2)$ (top panel) and $x F_{3A}^{\nu+\bar\nu}(x,Q^2)$ (bottom panel) vs $Q^2$ are shown at different values of $x$ in $^{56}$Fe  
and $^{208}$Pb. The results are obtained with the full model (solid line) at NNLO and are compared with the results of the available 
experimental data~\cite{Berge:1989hr, Oltman:1992pq, Tzanov:2005kr, Onengut:2005kv}.
Both the nuclear targets are treated as isoscalar.}
\label{fig2p}
\end{figure}

 For the (anti)neutrino scattering cross sections and structure functions, high statistics measurements have been done by
 CCFR~\cite{Oltman:1992pq}, CDHSW~\cite{Berge:1989hr} and NuTeV~\cite{Tzanov:2005kr} experiments in iron and by CHORUS~\cite{Onengut:2005kv} collaboration in lead nuclei.  These experiments have 
 been performed in a wide energy range, i.e., $20\le E_\nu \le 350$ GeV and measured the differential scattering cross sections. From these measurements the nuclear structure functions are extracted.
 We study the nuclear modifications for the (anti)neutrino induced processes in $F_{2A}^{\nu+\bar\nu}(x,Q^2)$ 
and $x F_{3A}^{\nu+\bar\nu}(x,Q^2)$ vs $Q^2$ in $^{56}$Fe and $^{208}$Pb nuclei by treating them as isoscalar nuclear targets.
The results are presented in Fig.\ref{fig2p} at the different values of $x$ using the full model at NNLO and are compared with the available
 experimental data from CCFR~\cite{Oltman:1992pq}, CDHSW~\cite{Berge:1989hr}, NuTeV~\cite{Tzanov:2005kr} and CHORUS~\cite{Onengut:2005kv} experiments. 
 We find a good agreement between the theoretical results for $F_{2A}^{\nu+\bar\nu}(x,Q^2)$ and reasonable agreement for $F_{3A}^{\nu+\bar\nu}(x,Q^2)$ with the experimental data.

\begin{figure}
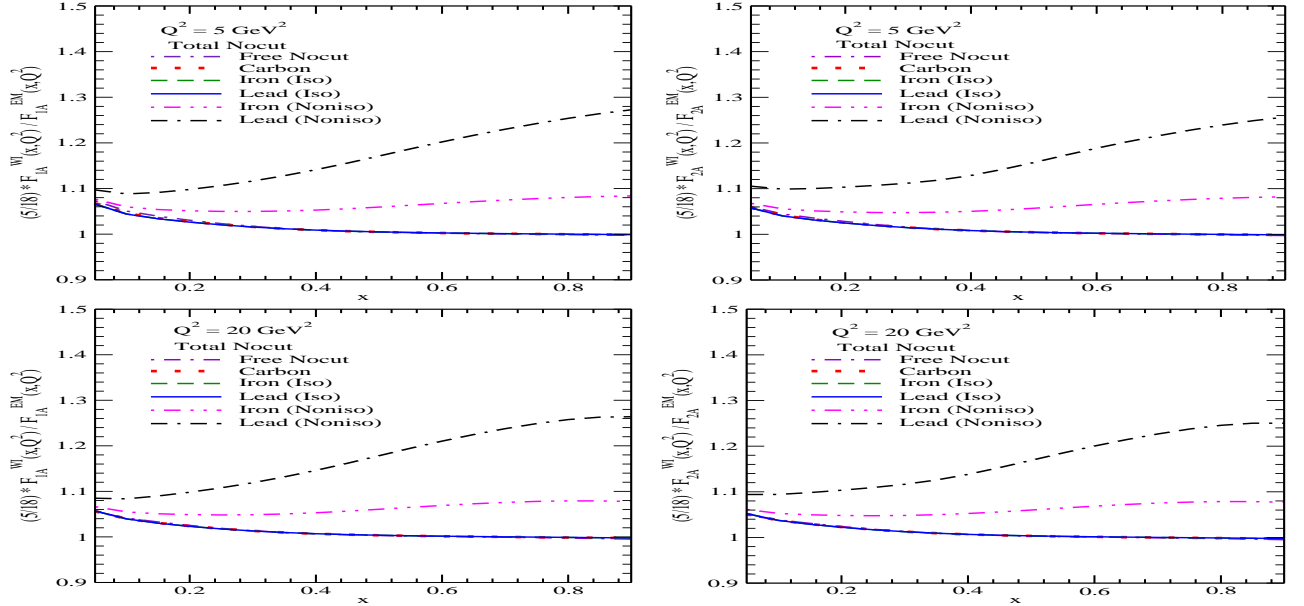

\begin{center}
 \includegraphics[height= 4.0 cm , width= 0.95\textwidth]{./fig/ratio_em_weak.eps}\\
  \includegraphics[height= 4.0 cm , width= 0.95\textwidth]{./fig/ratio_em_weak_20q2.eps}
\end{center}
 \caption{The ratio $R^{WI/EM}(x,Q^2)=\left(\frac{5}{18}\right)\;\frac{F_{iA}^{WI}(x,Q^2)}{F_{iA}^{EM}(x,Q^2)};~i=1,2$ vs $x$ are shown at $Q^2=5$ GeV$^2$ (top panel) and 20 GeV$^2$ (bottom panel) in $^{12}$C,
 $^{56}$Fe and $^{208}$Pb. The numerical results are obtained at NNLO using the full model and are compared with the free nucleon case.}
\label{fig7}
\end{figure}

We have also studied the nuclear modifications in the electromagnetic structure functions~\cite{Zaidi:2019mfd} and compared them with the weak structure functions
for the free nucleon target, isoscalar nuclear targets and nonisoscalar nuclear targets and present the results in Fig.\ref{fig7}
for the ratios $\left(\frac{5}{18}\right)\;\frac{F_{1A}^{WI}(x,Q^2)}{F_{1A}^{EM}(x,Q^2)}$ (left panel)
and $\left(\frac{5}{18}\right)\;\frac{F_{2A}^{WI}(x,Q^2)}{F_{2A}^{EM}(x,Q^2)}$ (right panel) vs $x$ at $Q^2=$ 5 and 20 GeV$^2$. The numerical results are shown at NNLO for carbon, iron and lead with 
the full model and are compared with the results of free nucleon. 
 It may be noticed from the figure that the ratio $R^{WI/EM}(x,Q^2)$ deviates from unity in the region of 
low $x$ even for the free nucleon case. It implies non-zero contribution from strange and charm quarks distributions which 
are found to be different in the case of electromagnetic and weak structure functions. However, for $x\ge 0.4$, where 
the contribution of strange and charm quarks are almost negligible, the ratio approaches $\sim 1$. Furthermore, if one assumes $s=\bar s$ and $c=\bar c$ then
in the region of small $x$, this ratio would be unity for an isoscalar nucleon target following the $\left(\frac{5}{18}\right)^{th}$-sum rule.
 It may be seen that the difference between the ratio $R^{WI/EM}(x,Q^2)$ for the isoscalar nuclear targets and the free nucleon target is almost negligible. 
The evaluation is also done for the nonisoscalar nuclear targets ($N>>Z$) like iron and lead. We must emphasize that in the present model, the spectral 
functions are normalized separately for the proton ($Z$) and neutron ($N=A-Z$) numbers in a nuclear target and to the number of 
nucleons for an isoscalar nuclear target~\cite{Haider:2015vea}. The ratio $R^{WI/EM}(x,Q^2)$ shows a significant 
deviation for the nonisoscalar nuclear targets which increases with nonisoscalarity, i.e. $\delta=\frac{(A-Z)}{Z}$. 
  This shows that the charm and strange quark distributions are 
  significantly different in asymmetric heavy nuclei as compared to the free nucleons.
It is important to notice that although some deviation is present in the entire range of $x$, it becomes more pronounced with the increase in $x$.
For example, in iron (nonisoscalar) the deviation from the free nucleon case is $2\%$ at $x=0.2$, $5\%$ at $x=0.5$, and 
$8\%$ at $x=0.8$ while in lead (nonisoscalar) it is found to be $\sim 7\%$ at $x=0.2$, $16\%$ at $x=0.5$, and $25\%$ at $x=0.8$ at $Q^2=5$ GeV$^2$. This deviation also has some
$Q^2$ dependence and with the increase in $Q^2$ the deviation becomes smaller. From the figure, it may be observed that the isoscalarity corrections, significant in the region of large $x$,
  are different in $F_{1A}(x,Q^2)$ and $F_{2A}(x,Q^2)$ albeit the difference is small. 

\begin{figure}
\begin{center}
 \includegraphics[height= 8 cm , width= 0.95\textwidth]{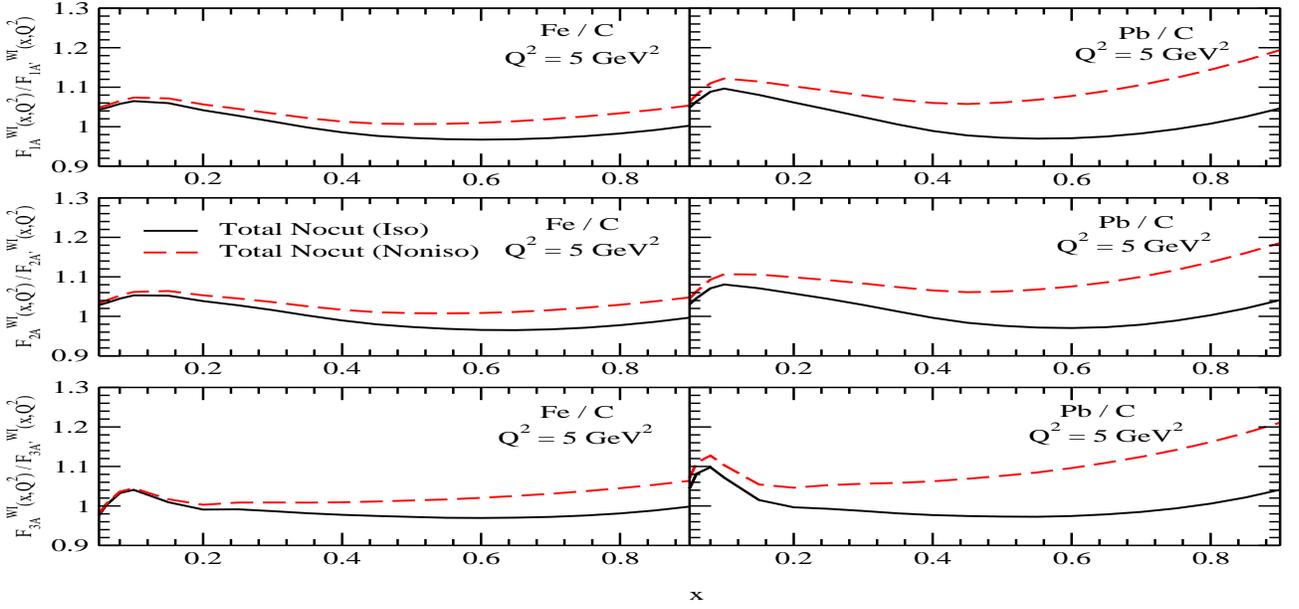}
\end{center}
\caption{$\frac{F_{iA}^{WI}(x,Q^2)}{F_{iA'}^{WI}(x,Q^2)};~(i=1-3$; $A=^{56}$Fe, $^{208}$Pb; $A'=^{12}$C) vs $x$ are shown at $Q^2=5$ GeV$^2$. The results are obtained using the full 
model at NNLO by 
treating $^{56}$Fe, $^{208}$Pb to be isoscalar (solid line) as well as nonisoscalar (dashed line) nuclear targets.}
\label{fig8}
\end{figure}
 We have also presented the results for the ratios of nuclear structure functions such as 
 \begin{equation}
   \frac{F_{iA}^{WI}(x,Q^2)}{F_{iA'}^{WI}(x,Q^2)};~(i=1,2,3;~A= ^{56}Fe,~ 
 ^{208}Pb~ \text{and}~ A'=^{12}C)~ \text{vs}~ x
 \end{equation} 
 at $Q^2=5$ GeV$^2$ in Fig.\ref{fig8}. The numerical results are shown with the full model at NNLO by treating iron and lead to be isoscalar as well as nonisoscalar nuclear targets. 
The following aspects are evident from the observation of Fig.\ref{fig8}:
 \begin{itemize}
  \item The deviation of the ratios $\frac{F_{i Fe}^{WI}(x,Q^2)}{F_{i C}^{WI}(x,Q^2)}$ and $\frac{F_{i Pb}^{WI}(x,Q^2)}{F_{i C}^{WI}(x,Q^2)}$ from unity in the 
  entire range of $x$ implies that nuclear medium effects are $A$ dependent. From the figure, it may also be noticed that the ratio in lead is higher than the ratio in iron
  which shows that medium effects become more pronounced with the increase in the nuclear mass number.
   There is noticeable enhancement in the ratio obtained for the nonisoscalar case from the results obtained for the isoscalar nuclear targets specially at high $x$. 
  This implies that nonisoscalarity effect increases with the increase in $x$
  as well as in the mass number.
  \item It is important to notice that although the behaviour of the ratio is qualitatively same in $\frac{F_{iA}^{WI}(x,Q^2)}{F_{iA'}^{WI}(x,Q^2)};~(i=1-3)$, quantitatively it is different. 
 \end{itemize}
  \begin{figure}
\begin{center}
 \includegraphics[height= 8 cm , width= 0.95\textwidth]{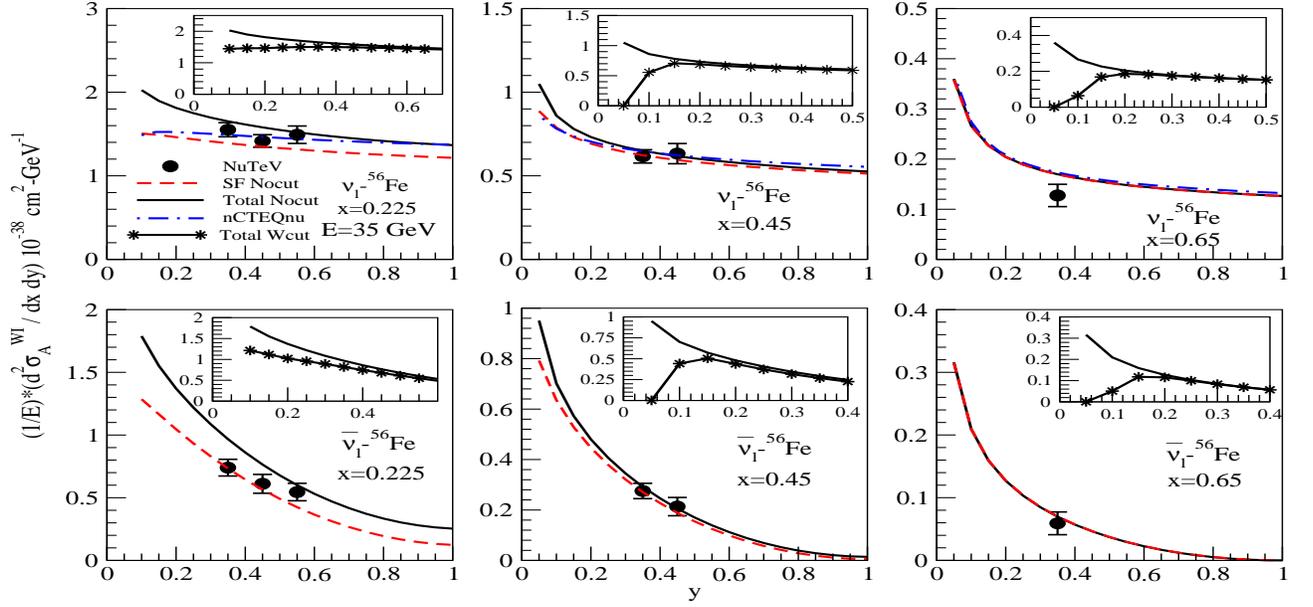}
\end{center}
 \caption{$\frac{1}{E}\;\frac{d^2\sigma_A^{WI}}{dx dy}$ vs $y$ are shown at the different values of $x$ for $E=35$ GeV. The numerical results for $\nu_l-^{56}$Fe DIS (top panel) and 
 $\bar\nu_l-^{56}$Fe DIS (bottom panel) processes are obtained with the spectral function only (dashed line) and with the full model (solid line) at NNLO. In the inset the results for the full model 
 are compared with the corresponding results obtained with a kinematical cut of $W>2$ GeV (solid line with star). Solid circles are the experimental data points 
 of NuTeV~\cite{Tzanov:2005kr}. The results for the $\nu_l$ induced process obtained using the nuclear PDFs nCTEQnu~\cite{private} (dash-dotted line) are also shown.}
\label{fig4a}
\end{figure}

\begin{figure}
\begin{center}
  \includegraphics[height= 8 cm , width= 0.95\textwidth]{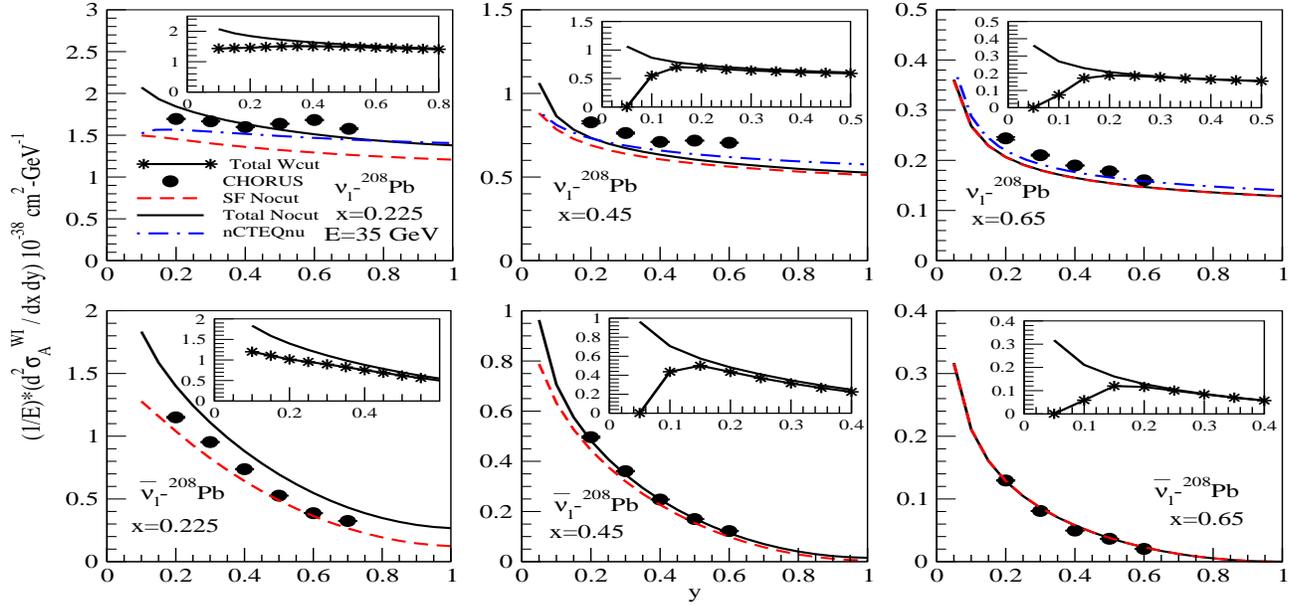}
\end{center}
 \caption{$\frac{1}{E}\;\frac{d^2\sigma_A^{WI}}{dx dy}$ vs $y$ are shown at the different values of $x$ for $E=35$ GeV. The numerical results for $\nu_l-^{208}$Pb DIS (top panel) and 
 $\bar\nu_l-^{208}$Pb DIS (bottom panel) processes are obtained with the spectral function only (dashed line) and with the full model (solid line) 
 at NNLO. In the inset the results for the full model 
 are compared with the corresponding results obtained with a kinematical cut of $W>2$ GeV (solid line with star). Solid circles are the experimental data points of 
 CHORUS~\cite{Onengut:2005kv}. The results for the $\nu_l$ induced process obtained using the nuclear PDFs nCTEQnu~\cite{private} (dash-dotted line) are also shown.}
\label{fig4b}
\end{figure}

\begin{figure}
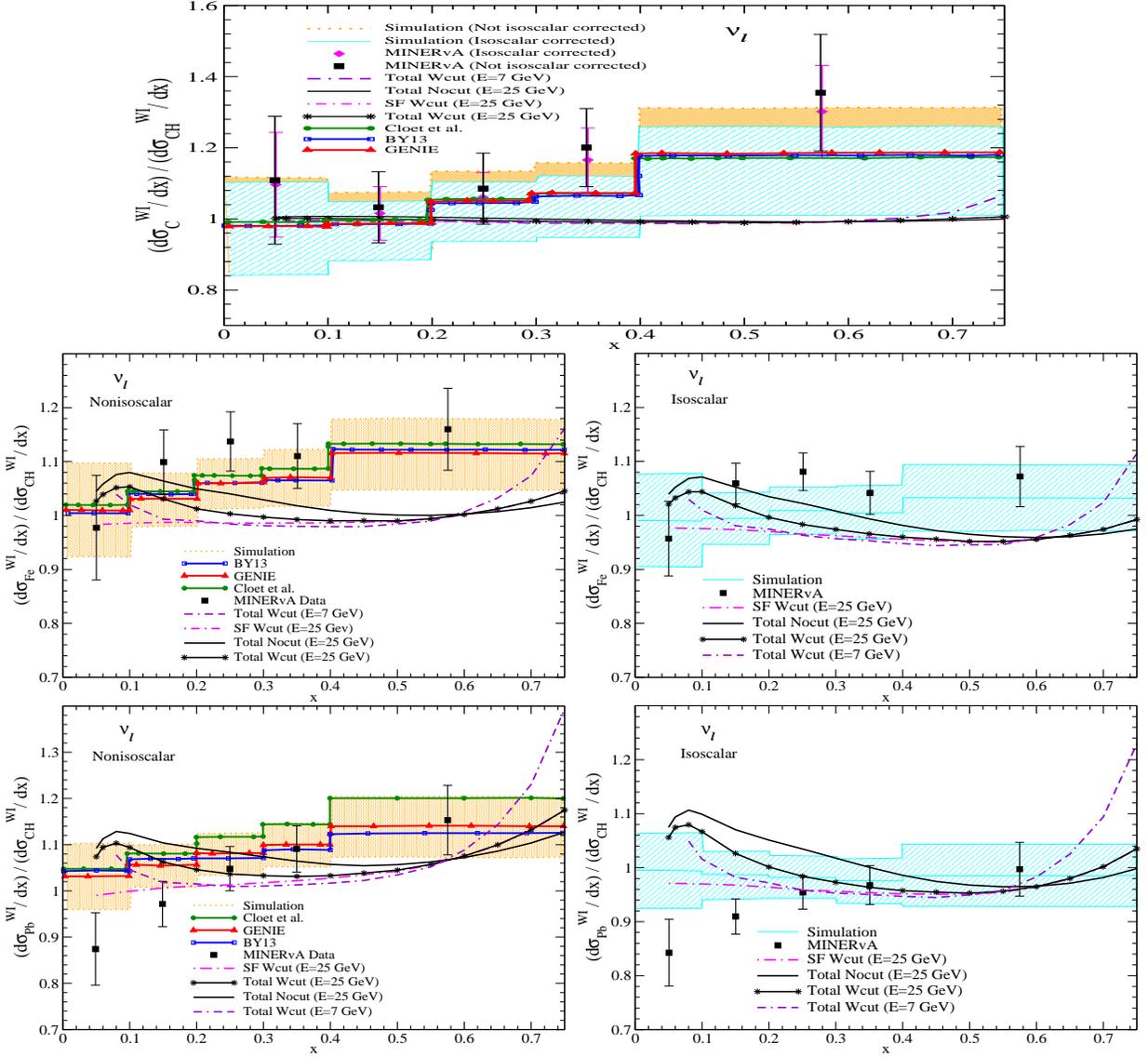

\begin{center}
  \includegraphics[height= 5. cm , width= 0.7\textwidth]{./fig/dsigma_ctoch_mix.eps}\\
  \includegraphics[height= 5. cm , width= 0.45\textwidth]{./fig/dsigma_fetoch.eps}
    \includegraphics[height= 5. cm , width= 0.45\textwidth]{./fig/dsigma_fetoch_isoscalar.eps}\\
  \includegraphics[height= 5. cm , width= 0.45\textwidth]{./fig/dsigma_pbtoch.eps}
    \includegraphics[height= 5. cm , width= 0.45\textwidth]{./fig/dsigma_pbtoch_isoscalar.eps}
\end{center}
 \caption{$\frac{d\sigma^{WI}_{A}/dx}{d\sigma^{WI}_{CH}/dx}~(A=$ $^{12}$C, $^{56}$Fe, $^{208}$Pb) vs $x$ for incoming neutrino beam of energies $E=7$ GeV and $25$ GeV. The numerical results are 
 obtained with the spectral function only (dash-dotted line: $E=25$ GeV) as well as with the full model (solid line: $E=25$ GeV, solid line with star: $E=25$ GeV
 and double dash-dotted line: $E=7$ GeV) at NNLO and are compared with the phenomenological results of Cloet et al.~\cite{Cloet:2006bq}, Bodek-Yang~\cite{Bodek:2010km}, 
 GENIE Monte Carlo~\cite{Andreopoulos:2009rq} and with the simulated results~\cite{Mousseau:2016snl}. The solid squares are the experimental points of MINERvA~\cite{Mousseau:2016snl}.
 The results in the left and right panels are respectively shown for the nonisoscalar and isoscalar nuclear targets.}
\label{fig6}
\end{figure}

In the literature the choice of a sharp kinematical cut on $W$ and $Q^2$ required to separate the regions of 
 nucleon resonances and DIS, i.e. region of shallow inelastic scattering and deep inelastic scattering is debatable. However, in some of the analysis the kinematic region of $Q^2>1$ GeV$^2$ and 
 $W>2$ GeV is considered to be the region of safe DIS~\cite{Mousseau:2016snl, Morfin:2012kn} and this has been taken into account in the analysis of
 MINERvA experiment~\cite{Mousseau:2016snl}. Therefore, to explore
 the transition region of nucleon resonances and DIS we have also studied the 
 effect of CoM cut on the scattering cross section. 
In Figs.~\ref{fig4a}-\ref{fig9}, we have presented the results with a 
  CoM cut of 2 GeV ($W>2$ GeV) and $Q^2>1$ GeV$^2$ which are labeled as ``Wcut'' and compared them with the corresponding ``Nocut'' results ($Q^2>1$ GeV$^2$ only)
  as well as with the available experimental data.

 In Figs.\ref{fig4a} and \ref{fig4b}, the results are shown for $\frac{1}{E}\;\frac{d^2\sigma_A^{WI}}{dx dy}$ vs $y$ for $\nu_l-A,~(A=^{56}Fe,~^{208}Pb)$ (top panel) and 
 $\bar\nu_l-A,~(A=^{56}Fe,~^{208}Pb)$ (bottom panel) scattering processes at NNLO. The numerical 
 results are obtained for a beam energy of 35 GeV at the different values of $x$. In Fig.\ref{fig4a}, theoretical results are
 presented for the spectral function only (dashed line) and for the full model (solid line) without having any cut on the CoM energy in iron and are compared with
 the NuTeV experimental data~\cite{Tzanov:2005kr}. 
 It may be seen that due to the mesonic contribution the results with the full model are higher than the results with the spectral function at $x=0.225$, however, for $x\ge 0.45$, where
 mesonic contribution is suppressed the difference becomes small. For example, in $\nu_l-^{56}$Fe($\bar\nu_l-^{56}$Fe) this enhancement is found to be $24\%(30\%)$ at $x=0.225$ and 
 $6\%(8\%)$ at $x=0.45$ for $y=0.2$. 
 Furthermore, we have compared these results with the phenomenological results of nCTEQnu~\cite{private} (evaluated by using$\nu_l - A$ scattering experimental data).
 One may notice that the present theoretical results differ from the results of nCTEQnu PDFs parameterization~\cite{private} in the region of low $x$ and $y$ while at high $x$ and $y$ 
they are in good agreement. 
In the inset of this figure, the results obtained with the full model having no cut on $W$ (solid line)
 are compared with the results obtained with a cut of $W>2$ GeV (solid line with star). It is 
 important to notice that the difference between these results becomes more pronounced with the increase in $x$ specially at low y, for example, at $y=0.1(y=0.4)$ there
 is a difference of $30\%(7\%)$ at $x=0.225$ and $36\%(3\%)$ at $x=0.45$ in $\nu_l-^{56}$Fe scattering process while for 
 $\bar\nu_l-^{56}$Fe it is found to be $32\%(13\%)$ and $37\%(8\%)$ respectively at $x=0.225$ and $x=0.45$. For higher values of $y$ the effect of CoM energy cut is small.
 However, there is no experimental data in the region of low $y$ to test these results.
 
 In Fig.\ref{fig4b}, we have presented the numerical results of the differential scattering cross section in $^{208}$Pb for the neutrino and antineutrino induced
 processes and compared them with the experimental data of
 CHORUS~\cite{Onengut:2005kv} experiment, where a comparison of the theoretical results for $\nu_l-^{208}Pb$ scattering has also been made 
 with the results of nCTEQnu~\cite{private} nuclear PDFs parameterization. We find that due to the $A$ dependence, the nuclear medium
 effects are more pronounced in lead as compared to iron and the effect of CoM energy cut causes
 relatively larger suppression in the region of low $x$ and $y(\le0.4)$. For 
 the numerical results presented in Figs.\ref{fig4a} and \ref{fig4b}, the nuclear targets are treated to be isoscalar.

The MINERvA experiment has used the NuMI neutrino beam at Fermilab for the cross section measurements in the low and medium energy modes that peak around neutrino
energy of 3 GeV and 6 GeV, respectively. 
The low energy neutrino broad band energy spectrum that peaks at $\sim 3$ GeV extends up to 100 GeV, however, neutrino flux drops steeply at 
high energies. MINERvA collaboration~\cite{Mousseau:2016snl} has reported the 
ratio of flux integrated differential scattering cross sections in carbon, iron and lead to the polystyrene scintillator ($CH$) vs $x$
in the neutrino energy range of 5-50 GeV. 
We have chosen two neutrino beam energies viz. $E=7$ GeV and 25 GeV, in a wide energy spectrum ($7 \le E \le 25$ GeV), in order to study the energy dependence of the nuclear medium effects. 
We have obtained $\frac{d\sigma^{WI}_A}{dx}$ by integrating Eq.\ref{d2sigdxdy_weak1} over $y$ in the limits 0 and 1 and present the theoretical results for the ratio 
$\frac{d\sigma^{WI}_{A}/dx}{d\sigma^{WI}_{CH}/dx}~(A=^{12}C, ^{56}Fe, ^{208}Pb)$ at $E=7$ GeV and 25 GeV for the charged current $\nu_l-A$ and $\bar\nu_l-A$ DIS processes.
The theoretical results are obtained in the kinematic region relevant for the MINERvA experiment ($W>2$ GeV and $Q^2>1$ GeV$^2$) and compared with the experimental data as well as with the 
results obtained using phenomenological models of Cloet et al.~\cite{Cloet:2006bq}, Bodek-Yang~\cite{Bodek:2010km} and GENIE Monte Carlo~\cite{Andreopoulos:2009rq}.

The results for the ratio ($\frac{d\sigma^{WI}_{A}/dx}{d\sigma^{WI}_{CH}/dx}$) vs $x$ in the case of $\nu_l-A$ scattering are presented in Fig.\ref{fig6} and are summarized below:
\begin{itemize}
 \item As the nuclear medium effects are approximately the same in carbon and $CH$, therefore, the ratio $\frac{d\sigma^{WI}_{C}/dx}{d\sigma^{WI}_{CH}/dx}$ (top panel) is expected
 to be close to unity.
 From the Fig.\ref{fig6}, one may notice that the deviation of the ratio from unity is 
 small in $ \frac{d\sigma^{WI}_{C}/dx}{d\sigma^{WI}_{CH}/dx}$, however, for $\frac{d\sigma^{WI}_{Fe}/dx}{d\sigma^{WI}_{CH}/dx}$ and $\frac{d\sigma^{WI}_{Pb}/dx}{d\sigma^{WI}_{CH}/dx}$ 
 it becomes large which shows the $A$ dependence of the nuclear medium effects specially the contribution of mesons which increases with $A$ at low and intermediate $x$. For example, at $E=25$ GeV 
 the contribution of mesons is found to be $10\%(7\%)$ at $x=0.1$, $2\%(1\%)$ at $x=0.3$, and $<1\%$ at
 $x=0.6$ in lead(iron) when they are treated to be isoscalar. It is important to notice that even for high energy neutrino beams 
  the effect of nuclear medium on the differential scattering cross section are significant.
  
 \item  We have found that due to the mass dependence of nuclear medium effects, the difference between the results of $\frac{d\sigma^{WI}_{C}/dx}{d\sigma^{WI}_{CH}/dx}$ 
 and $\frac{d\sigma^{WI}_{Fe}/dx}{d\sigma^{WI}_{CH}/dx}\left(\frac{d\sigma^{WI}_{Pb}/dx}{d\sigma^{WI}_{CH}/dx}\right)$
 obtained by using the full model at $E=25$ GeV (solid line) is $\simeq 4\%(7\%)$ at $x=0.05$, $6\%(9\%)$ at $x=0.1$ and $3\%(\sim 3\%)$ at $x=0.6$ when there is no constrain on the CoM energy $W$. 
 While the cut of $W>2$ GeV, leads to a change of $1-5\%$ in this 
 difference in the entire range of $x$, for example, there is further reduction of $\simeq 2\%$ at $x=0.05$, $3\%$
 at $x=0.1$, $\simeq 5\%$ at $x=0.2$ and $<1\%$ at $x=0.6$ in the differential scattering cross section.  
 
  \item To study the isoscalarity effect we have obtained the results for $\frac{d\sigma^{WI}_{Fe}/dx}{d\sigma^{WI}_{CH}/dx}$ and $\frac{d\sigma^{WI}_{Pb}/dx}{d\sigma^{WI}_{CH}/dx}$ 
 by treating iron and lead to be nonisoscalar (left panel) as well as isoscalar (right panel) targets (Fig.\ref{fig6}). The isoscalarity correction in asymmetric nucleus
 is found to be significant. For example, at $E= 25$ GeV, this effect is $2\%(5\%)$, and $5\%(13\%)$ at $x=0.3$ and $0.7$, respectively, in iron(lead) when no kinematical cut is applied on $W$.

 \item To observe the energy dependence of the scattering cross section, numerical results obtained using the full model with $Q^2>1$ GeV$^2$ and $W>2$ GeV at
 $E=25$ GeV (solid line with star) are compared 
with the corresponding results obtained at $E=7$ GeV (double dash-dotted line). It may be observed that in the region of low and intermediate $x$ 
the results for $ \frac{d\sigma^{WI}_{A}/dx}{d\sigma^{WI}_{CH}/dx}$ at $E=7$ GeV are smaller in magnitude from the results at $E=25$ GeV while with the increase
in $x$ the ratio $ \frac{d\sigma^{WI}_{A}/dx}{d\sigma^{WI}_{CH}/dx}$ obtained for $E=7$ GeV increases. Due to the energy dependence of the 
  differential scattering cross section, the difference between the results obtained using the full model at the aforesaid energies, i.e. 7 GeV and 25 GeV,
  is $\simeq 3\%(5\%)$, $ 2\%(\simeq 2\%)$ and $12\%(\simeq 16\%)$ at
 $x=0.1$, $x=0.3$ and $x=0.75$ , respectively, if iron (lead) is treated as isoscalar nuclear target. 
\begin{figure}[hb]
\begin{center}
  \includegraphics[height= 15 cm , width= 0.90\textwidth]{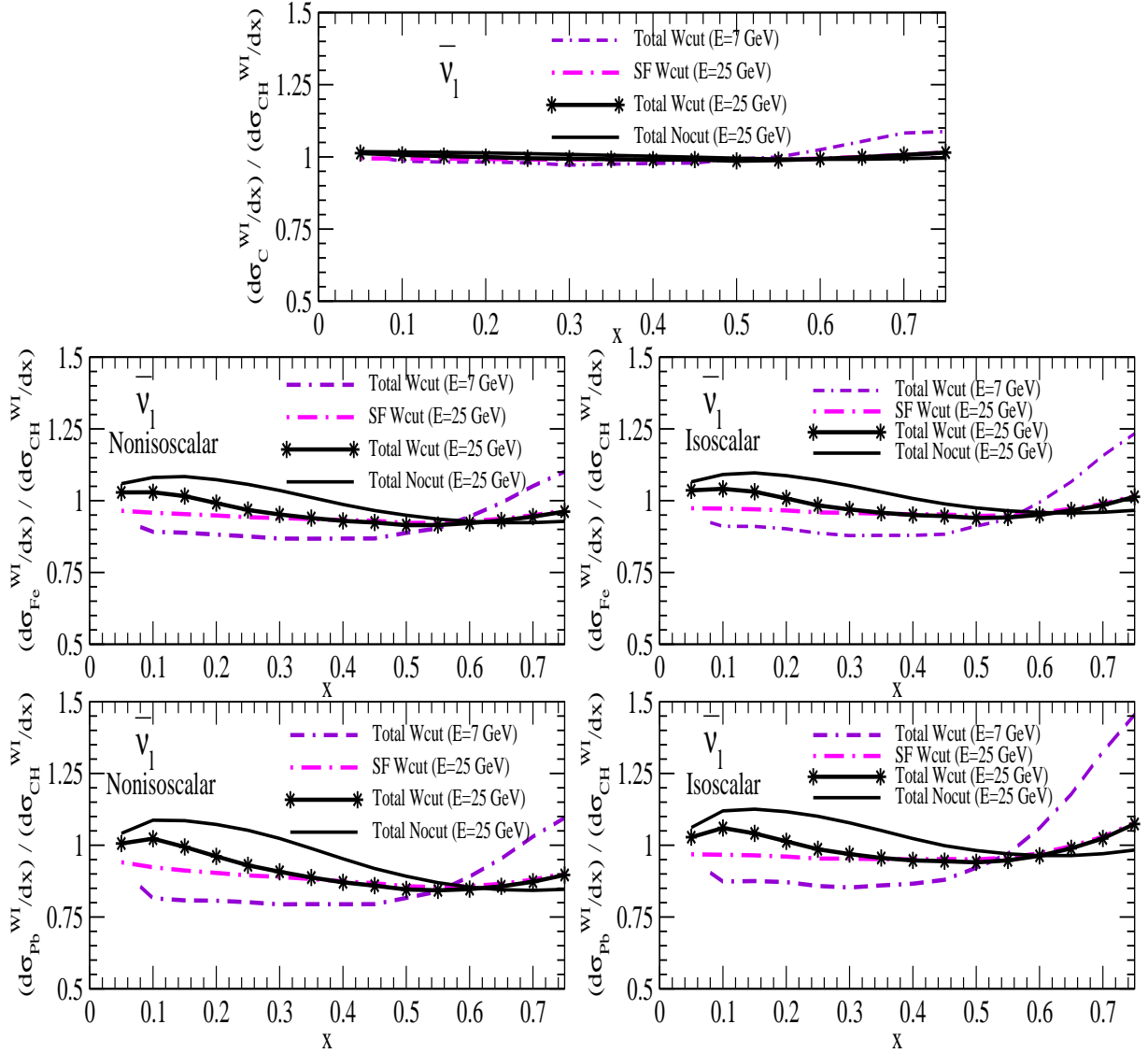}
\end{center}
 \caption{$\frac{d\sigma^{WI}_{A}/dx}{d\sigma^{WI}_{CH}/dx}~(A=^{12}C, ^{56}Fe, ^{208}Pb)$ vs $x$ are shown for incoming antineutrino beam of energies $E=7$ GeV and $25$ GeV at NNLO. 
 The numerical results are obtained using the spectral
 function only (dash-dotted line: $E=25$ GeV) by applying a cut of $W>2$ GeV and $Q^2>1$ GeV$^2$. The results using the full model are obtained with (solid line with star: $E=25$ GeV
 and double dash-dotted line: $E=7$ GeV) and without (solid line: $E=25$ GeV) a
 cut of 2 GeV on the CoM energy but keeping $Q^2>1$ GeV$^2$. The results in the left and right panels are respectively shown for the nonisoscalar and isoscalar nuclear targets. }
\label{fig9}
\end{figure}
 \item Furthermore, we have compared our theoretical results with the corresponding experimental data of MINERvA as well as with the different phenomenological models like that of 
 Cloet et al.~\cite{Cloet:2006bq} (solid line with circle), Bodek et al.~\cite{Bodek:2010km} (solid line with square)
 and GENIE MC~\cite{Andreopoulos:2009rq} (solid line with triangle). It may be noticed that MINERvA's experimental data have large error bars due to statistical uncertainties
 and the wide band around the simulation is due to the systematic errors which shows an uncertainty up to $\sim 20\%$~\cite{Mousseau:2016snl}. Although the results of phenomenological models lie in this 
 systematic error band even then none of the phenomenological model is able to describe the observed ratios in the whole region of $x$. 
\end{itemize}

We have also made predictions for the $\bar\nu_l - A$ scattering cross sections in the same kinematic region as considered in Fig.\ref{fig6} corresponding to the MINERvA experiment
and presented the results in Fig.\ref{fig9}, for the ratio $\frac{d\sigma^{WI}_{A}/dx}{d\sigma^{WI}_{CH}/dx}; ~(A=^{12}C,~^{56}Fe,~^{208}Pb)$ vs $x$ 
at $E=$7 GeV and 25 GeV without and with a cut of $W>2$ GeV. The nuclear medium effects in  $\frac{d\sigma^{WI}_{A}}{dx}$ for $\bar\nu_l-A$  scattering are found to be qualitatively similar to 
  $\nu_l-A$ scattering when no cut on CoM energy is applied, however, quantitatively they are different specially at low and mid values of $x$. For example, at $E=7$ GeV the enhancement 
 in the cross section when full calculations is applied from the results obtained using the spectral function 
  is about 24$\%$ at $x=0.25$ in $\nu_l-^{208}Pb$ scattering, while it is 65$\%$ in ${\bar\nu}_l-^{208}Pb$ scattering, and the
 difference in the two results decreases with the increase in $x$. At $E=25$ GeV the enhancement 
 in the cross section is about 20$\%$ at $x=0.25$ in $\nu_l-^{208}Pb$ scattering, while it is $\sim$45$\%$ in ${\bar\nu}_l-^{208}Pb$ scattering.
 When a cut of 2 GeV is applied on the CoM energy, then a suppression in the region of low and mid $x$ is observed in the differential cross section, resulting in a lesser
 enhancement due to mesonic effects,
  for example, at $E=25$ GeV, the enhancement due to the 
 mesonic contributions becomes $\sim$18$\%$ (vs 20\% without cut) in ${\nu}_l-^{208}Pb$ scattering while 
 $\sim 28\%$ (vs 45\% without cut) in ${\bar\nu}_l-^{208}Pb$ scattering at $x=0.25$. At $E=7$ GeV, with a cut of 2 GeV on $W$, the enhancement is about 2$\%$ at $x=0.25$ 
 in $\nu_l-^{208}Pb$ scattering, 
 while there is reduction in $\bar\nu_l-A$  scattering, implying small contribution from the mesonic part. This reduction in $\frac{d\sigma^{WI}_{A}}{dx}$ for 
 $\bar\nu_l-A$ scattering is about 15$\%$ in a wide region of $x$($\le$ 0.6).
 When the results for $\frac{d\sigma^{WI}_{A}/dx}{d\sigma^{WI}_{CH}/dx}$ using antineutrino beam are compared with neutrino results, we find that without any cut on $W$, the results
 are similar, but with a cut for $E=$7 GeV there is enhancement at high $x$. This enhancement is larger
 in $^{208}Pb$ than in $^{56}Fe$ due to large effect of Fermi motion in heavy nuclei. 

\section{Summary and Conclusion}\label{summary}
Our findings for the weak nucleon and nuclear structure functions and the differential scattering cross sections are as follows:

 \begin{itemize}
\item  The difference in the results of free nucleon structure functions $F_{iN}^{WI}(x,Q^2)~(i=1,2)$ evaluated at NLO with HT effect and the results obtained at NNLO 
 is almost negligible ($<1\%$). However, this difference is somewhat larger for $F_{3N}^{WI}(x,Q^2)$ at low $x$ and low $Q^2$ which becomes small with the increase in $Q^2$.
 In the case of nucleons bound inside a nucleus, the HT corrections are further suppressed due to the presence of nuclear medium effects. Consequently, the results for 
 $\nu_l/\bar\nu_l-A$ DIS processes which are evaluated at NNLO have almost negligible difference from the results obtained at NLO with HT effect.
 \item The nuclear structure functions obtained with spectral function only is suppressed from the free nucleon case in the entire region of $x$.
 Whereas, the inclusion of mesonic contributions results in an enhancement in the nuclear structure functions in the low and intermediate region of $x$. 
 Mesonic contributions are observed to be more pronounced with the increase in mass number and they decrease with the increase in $x$ and $Q^2$.
The results for the nuclear structure functions $F_{2A}^{WI}(x,Q^2)$ and $F_{3A}^{WI}(x,Q^2)$ with the full theoretical model show good agreement with the 
 experimental data of CCFR~\cite{Oltman:1992pq}, CDHSW~\cite{Berge:1989hr}, NuTeV~\cite{Tzanov:2005kr} and CHORUS~\cite{Onengut:2005kv} especially at high $x$ and high $Q^2$. Predictions are 
 also made for $^{40}Ar$ that may be useful in analyzing the experimental results of DUNE~\cite{Abi:2018alz, Acciarri:2015uup} and ArgoNeuT~\cite{Acciarri:2014isz}.
 
 \item We have found nuclear medium effects to be different in electromagnetic and weak interaction channels specially for the 
 nonisoscalar nuclear targets. The contribution of strange and charm quarks is found to be different for the electromagnetic and weak interaction induced processes off free
 nucleon target which also gets modified differently for the heavy nuclear targets.
 Furthermore, we have observed that the isoscalarity corrections, significant even at high $Q^2$, are not the same in $F_{1A}^{WI}(x,Q^2)$ and $F_{2A}^{WI}(x,Q^2)$.
 
 \item The nuclear medium effects are found to be important in the evaluation of differential scattering cross section. We have observed that 
 in the $\bar\nu_l-A$ reaction channel the nuclear medium effects are more pronounced than in the case of $\nu_l-A$ scattering process.
 Our results of $\frac{1}{E}\;\frac{d^2\sigma_A^{WI}}{dx dy};$ $(A=^{56}Fe,~^{208}Pb)$ obtained using the full model show a reasonable agreement
with the experimental data of NuTeV~\cite{Tzanov:2005kr} and CHORUS~\cite{Onengut:2005kv} for the neutrino and antineutrino induced DIS processes.
 Theoretical results of differential cross section are also found to be in good agreement with the phenomenological results 
of nCTEQnu nuclear PDFs parameterization~\cite{private} in the intermediate as well as high region of $x$ for all values of $y$. 
 
\item The present theoretical results for the ratio $\frac{d\sigma_A^{WI}/dx}{d\sigma_{CH}^{WI}/dx}~(A=^{12}C,~^{56}Fe,~^{208}Pb)$ when
compared with the different phenomenological models and MINERvA's experimental data on $\nu_l-A$ scattering, imply that a better understanding of nuclear medium effects is
required in the $\nu_l(\bar\nu_l)-$nucleus deep inelastic scattering. We have also
made predictions for the $\bar\nu_l-A$ DIS cross sections relevant for the upcoming MINERvA results. 

\end{itemize}

To conclude, the present theoretical results provide information about the energy dependence, effect of CoM energy cut, 
medium modifications and isoscalarity correction effects on the nuclear structure functions and cross sections for the deep inelastic
scattering of (anti)neutrino from various nuclei. This study will be helpful
to understand the present and future experimental results from MINERvA~\cite{Mousseau:2016snl}, ArgoNeuT~\cite{Acciarri:2014isz}, and DUNE~\cite{Abi:2018alz, Acciarri:2015uup} experiments.

\section*{Acknowledgment}   
F. Zaidi is thankful to the Council of Scientific \& Industrial Research (CSIR), India, for providing the research associate fellowship with 
award letter no. 09/112(0622)2K19 EMR-I.
M. S. A. and S. K. S. are thankful to Department of Science and Technology (DST), Government of India for providing 
financial assistance under Grant No. EMR/2016/002285.  I.R.S. acknowledges support from Spanish Ministerio de Economia y Competitividad under grant No. FIS2017-85053-C2-1-P, and by Junta de Andalucia (Grant No. FQM-225).

\end{document}